\documentclass[ aip, jmp, preprint, showpacs,preprintnumbers,amsmath,amssymb]{revtex4-1}
\usepackage{amsmath}
\usepackage{graphicx}% Include figure files
\usepackage{dcolumn}% Align table columns on decimal point
\usepackage{bm}% bold math
\usepackage{color}

\definecolor{DarkGreen}{rgb}{0.0,0.45,0.0}     % define a custom color
\definecolor{DarkMagenta}{rgb}{0.45,0.0,0.45}  % define a custom color

\begin{document}

%\preprint{APS/123-QED}

\title{Magnetic reconnection in the low solar chromosphere with a more realistic radiative cooling model}% Force line breaks with \\

\author{Lei Ni }
\email{leini@ynao.ac.cn}
\affiliation{
Yunnan Observatories, Chinese Academy of Sciences, P. O. Box 110, Kunming, Yunnan 650216, P. R. China
 }
 \affiliation{
Center for Astronomical Mega-Science, Chinese Academy of Sciences, 20A Datun Road, Chaoyang District, Beijing 100012, P. R. China
 }
 
\author{Vyacheslav S. Lukin}
\affiliation{
National Science Foundation, 2415 Eisenhower Ave, Alexandria, VA 22314, USA
 }
 
 \author{Nicholas A. Murphy }
\affiliation{
Harvard-Smithsonian Center for Astrophysics, 60 Garden Street, Cambridge, Massachusetts 02138, USA
}
 
\author{Jun Lin}
\affiliation{
Yunnan Observatories, Chinese Academy of Sciences, P. O. Box 110, Kunming, Yunnan 650216, P. R. China
}
\affiliation{
Center for Astronomical Mega-Science, Chinese Academy of Sciences, 20A Datun Road, Chaoyang District, Beijing 100012, P. R. China
}

%\altaffiliation{Center for Integrated Computation and Analysis of Reconnection and Turbulence}
%\affiliation{

\date{\today}% It is always \today, today,
  

%\maketitle     
 
\begin{abstract}
Magnetic reconnection is the most likely mechanism responsible for the high temperature events that are observed in strongly magnetized locations around the temperature minimum in the low solar chromosphere. This work improves upon our previous work ["Magnetic Reconnection in Strongly Magnetized Regions of the Low Solar Chromosphere", The Astrophysical Journal 852, 95 (2018) ] by using a more realistic radiative cooling model computed from the OPACITY project and the CHIANTI database.  We find that the rate of ionization of the neutral component of the plasma is still faster than recombination within the current sheet region. For low $\beta$ plasmas, the ionized and neutral fluid flows are well-coupled throughout the reconnection region resembling the single-fluid Sweet-Parker model dynamics. Decoupling of the ion and neutral inflows appears in the higher $\beta$ case with $\beta_0=1.46$, which leads to a reconnection rate about three times faster than the rate predicted by the Sweet-Parker model.  In all cases, the plasma temperature increases with time inside the current sheet, and the maximum value is above $2\times10^4$~K when the reconnection magnetic field strength is greater than $500$~G. While the more realistic radiative cooling model does not result in qualitative changes of the characteristics of magnetic reconnection, it is necessary for studying the variations of the plasma temperature and ionization fraction inside current sheets in strongly magnetized regions of the low solar atmosphere. It is also important for studying energy conversion during the magnetic reconnection process when the hydrogen-dominated plasma approaches full ionization. 
\end{abstract}

%\pacs{plasma physics}% PACS, the Physics and Astrono                         % Classification Scheme.
\maketitle

\section{Introduction}
The low solar atmosphere host numerous small-scale events related to magnetic reconnection \cite{2015ApJ...798L..11Y, 2016NatCo...711837X, 2017ApJ...836...52Z}. These events have been observed by both ground based solar telescopes [e.g. the \emph{Swedish Solar Telescope}(SST), the \emph{New Vacuum Solar Telescope}(NVST), the \emph{Goode Solar Telescope}(GST)] as well as space-based observatories [e.g., the \emph{Solar Dynamics Observatory} (\emph{SDO}), the \emph{Interface Region Imaging Spectrograph} (\emph{IRIS})]. The most interesting and famous activities are the Ellerman Bomb type events \cite{2014Sci...346C.315P, 2015ApJ...812...11V, 2016A&A...593A..32G, 2016ApJ...824...96T, 2018APJ}, which are usually called IRIS bombs on UV bursts. 

The traditional Ellerman bomb (EB) is a transient and prominent enhancement in brightness of the far wings of spectral lines, in particular the Balmer H${\alpha}$ line at the wavelength of 6563\r{A}. An EB is usually observed in the solar active region. In the original work of Ellerman\cite{1917ApJ....46..298E}, EBs were described as "a very brilliant and very narrow band extending four or five angstroms on either side of the H${\alpha}$ line, but not crossing it, "gradually disappearing in a few minutes. Ellerman (1917) originally named them "solar hydrogen bombs"; and nowadays, this brightening phenomenon is named after him. The EB takes place exclusively in the region of new emerging flux, and manifests characteristic elongated shaped when showing in high resolution images. Usually, the temperature of the region where as EB occurs is below $10^4$~K \cite{2006ApJ...643.1325F, 2014ApJ...792...13H, 2015ApJ...798...19N}. 

The EB-type events have been observed both in the wings of H${\alpha}$ and the IRIS Si IV passbands, and they share some characteristics with traditional EBs [e.g., similar locations ( $250-750$~km above the solar surface), similar life times (about 3-5 min) and sizes (about $0.3^{\prime \prime}$-$0.8^{\prime \prime}$)]. Both are considered to be formed in magnetic reconnection processes. However, the emission in the Si IV 139.3 nm line requires a temperature of at least $2\times10^4$~K in the dense photosphere. Therefore, the maximum temperatures inside these EB-type events are much higher than traditional EBs. 

Most recently, Tian et al. (2018) have discovered the inverted Y-shape prevalent jets from sunspot light bridges \cite{2018APJ} by using multi-wavelength observations from both GST and IRIS. The transient brightenings with significant heating and bi-directional flows at the foot points of these jets indicated that magnetic reconnection drove the formation of the inverted Y-shape jets. These transient brightenings are also the EB-type events where the presence of Fe II and Ni II absorption lines indicated that the hot reconnection region was located below the cooler chromosphere. Additionally, the O IV 1401.16\r{A} and 1399.77\r{A} forbidden lines were almost absent during these compact brightenings, which demonstrated that the reconnection site was in a region of very high plasma density\cite{2014Sci...346C.315P}. 

%Therefore, a good reconnection model and theory is urgently needed to explain the characteristics of these Ellerman Bomb type events.

Previous numerical simulations have studied the formation of EBs through magnetic reconnection by using the one-fluid MHD model \cite{2001ChJAA...1..176C, 2007ApJ...657L..53I, 2009A&A...508.1469A, 2011RAA....11..225X}. The maximum temperature increase observed in these simulations modeling the conditions between the photosphere and the low solar chromosphere was no more than several thousand Kelvin. However, the resolutions in these simulations were low, the magnetic diffusion in the reconnection region was not realistic, and the important interactions between ions and neutrals were ignored. The high resolution simulations with a more realistic magnetic diffusion in Ni et al. (2016) showed that the plasma can be heated from $4200$~K to above $8 \times10^4$~K inside the multiple magnetic islands of a reconnection process with strong magnetic fields ($500$~G) in the temperature minimum region (TMR) of the solar atmosphere (about $500$~km above the solar surface). Ambipolar diffusion, temperature-dependent magnetic diffusion, heat conduction, and optically thin radiative cooling were all included \cite{2016ApJ...832..195N}. However, in this work the plasma was assumed to be in a steady ionization equilibrium state. Most recently, one-fluid 3D MHD simulations with radiative transport studied EBs and flares at the surface and in the lower atmosphere of the Sun \cite{2017ApJ...839...22H}. In these simulations, the plasma temperature was observed to remain below $10^4$~K during the EB formation process in the photosphere. However, non-equilibrium ionization effects were not considered in their model, and the artificial hyper-diffusivity operator that was included to prevent the collapse of the current sheets leaves open the possibility of smaller scale and hotter structures at spatial scales not covered in that simulation.

Background plasma near the TMR is weakly ionized and the plasma density is high. Thus, realistic simulations of magnetic reconnection in this region of the solar atmosphere must account for the interactions between ions and neutrals as well as radiative cooling. It has been shown that ambipolar diffusion which results from collisions between ions and neutrals makes current sheet thin rapidly when no guide field is present \cite{1994ApJ...427L..91B, 1995ApJ...448..734B, 1999ApJ...511..193V, 2015ApJ...799...79N}. Previous 1D analytical work also studied magnetic reconnection in weakly ionized plasma \cite{1999ApJ...511..193V, 2003ApJ...583..229H, 2004ApJ...603..180L}. They found that an excess of ions can build up in the reconnection region if the ions pulled in by the reconnecting magnetic field are decoupled from the neutrals. High recombination can then produce a loss of ions in the reconnection region that prevents ion pressure from building up further, which leads to faster magnetic reconnection independent of magnetic diffusivity. Leake et al. (2012, 2013) have used the reactive multi-fluid plasma-neutral module within the HiFi modeling framework to study null-point magnetic reconnection in the solar chromosphere \cite{2012ApJ...760..109L, 2013PhPl...20f1202L}. They showed that strong ion recombination in the reconnection region, combined with Alfv\'enic outflows, leads to faster reconnection within a two-dimensional (2D) numerical model. When strong radiative losses cool the plasma, it decreases the pressure inside the reconnection region and also results in the thinning of a current sheet when no guide field is present \cite{1995ApJ...449..777D, 2011PhPl...18d2105U}. Recently, Alvarez Laguna et al. (2017) studied how the radiative cooling affects magnetic reconnection in the solar chromosphere by using the same reactive multi-fluid plasma-neutral model as Leake et al. (2012) but with a different code \cite{2017ApJ...842..117A}. They found that the radiative losses strongly decreased the ion density and plasma pressure in a case with a high initial ionization fraction, which resulted in the rapid thinning of the current sheet and an enhancement of the reconnection rate \cite{2017ApJ...842..117A}. 

In all the previous reactive multi-fluid plasma-neutral simulations \cite{2012ApJ...760..109L, 2013PhPl...20f1202L, 2015ApJ...805..134M, 2017ApJ...842..117A}, the background neutral density ($3\times10^{18}-8\times10^{18}$~m$^{-3}$) is representative of a plasma above the middle chromosphere ($1000-1500$~km above the solar surface) according to the VAL-C solar atmosphere model \cite{1981ApJS...45..635V}. The magnetic field strength is only around $10$~G. The plasma $\beta$ in each of these simulations is above 1. However, in order to study magnetic reconnection in EB-type events, a different choice of the background plasma and neutral density to be characteristic of the TMR, i.e., approximately two orders of magnitude higher than those used in the previous papers, has to be made. Further, the EB-type events described by Tian et al. (2018) were observed in a sunspot region with the maximum bi-directional flow speed as high as $200$~km\,s$^{-1}$ \cite{2018APJ}. The reconnection magnetic field in these events may therefore be higher than $1000$~G. 

Our recent numerical simulations \cite{2018ApJ...852...95N} have studied magnetic reconnection in strongly magnetized regions around the solar TMR. In that paper, we have presented the first reactive multi-fluid simulations of magnetic reconnection in low $\beta$ plasmas with a guide field. The simulation results were significantly different from the previous high $\beta$ simulations with zero guide field. We found that the neutrals and ions were well-coupled throughout the reconnection region for the low $\beta$ plasma. The neutral and ionized fluid components decoupled upstream of the reconnection site only when the plasma $\beta$ was sufficiently high. The rate of ionization of the neutral component of the plasma was always faster than recombination within the current sheet region; and the initially weakly ionized plasmas could become fully ionized within the reconnection region when plasma $\beta$ was low enough. The current sheet could be strongly heated to high temperatures (above $2.5 \times10^4$~K) only when the reconnecting magnetic field was in excess of a kilogauss and the plasma inside became fully ionized. However, only a simple radiative cooling model \cite{2012ApJ...760..109L, 2013PhPl...20f1202L, 2018ApJ...852...95N} was applied in Ni et al. (2018). In particular, this simple model neglects the presence of minority species in the low solar atmosphere and can significantly underestimate radiative losses in a hydrogen-dominated plasma approaching full ionization.

In this work, we use a more realistic radiative cooling model \cite{2017ApJ...842..117A} to simulate magnetic reconnection around the solar TMR. This stronger radiative cooling model may be expected to result in faster recombination than ionization and to significantly impact the magnetic reconnection process as shown in Alvarez Laguna et al. (2017)\cite{2017ApJ...842..117A}. Including such a strong radiative cooling model may also be expected to reduce the temperature increases during the reconnection process, such that the high temperature plasma (above $2.5 \times10^4$~K) observed in our previous work would not appear even for magnetic fields in excess of a kilogauss. Therefore, the numerical results and conclusions in this work were expected to be significantly different from those reported in Ni et al. (2018)\cite{2018ApJ...852...95N}. Section II describes our numerical model and simulation setup.  We present our numerical results and compare them with our previous work \cite{2018ApJ...852...95N} in Section III.  A summary and discussion are given in Section IV.

\section{Normalization and Initial Conditions}

Our simulations are performed by using the reactive multi-fluid plasma-neutral module of the HiFi modeling framework\cite{2015ApJ...805..134M, 2018ApJ...852...95N}. Here, we normalize the equations by using the characteristic plasma density and magnetic field around the solar TMR. The characteristic values set in our simulations are exactly the same as in Ni et al. (2018), the characteristic plasma number density is $n_{\star}=10^{21}$~m$^{-3}$, and the characteristic magnetic field is $B_{\star}=0.05$~T=$500$~G, the characteristic length of $L_{\star}=100$~m. We have also derived the additional normalizing values to be $V_{\star}\equiv B_{\star}/\sqrt{\mu_0m_pn_{\star}}=34.613$~km\,s$^{-1}$, $t_{\star}\equiv L_{\star}/V_{\star}=t_A=0.0029$~s and $T_{\star} \equiv B_{\star}^2/(k_B\mu_0n_{\star})=1.441\times10^5$~K. The initial ionized and neutral fluid densities are set to be uniform with a neutral particle number density of $n_{n0}=0.5n_{\star}=0.5\times10^{21}$~m$^{-3}$, and the initial ionization degree is $f_{i0}=n_{i0}/(n_{i0}+n_{n0})=0.01\%$. Thus, the neutral-ion collisional mean free path of the background plasma is $\lambda_{ni0}=23.74$~m by assuming the cross-section $\Sigma_{ni}=5\times10^{-19}$ m$^{2}$. The ion inertial length is $d_{i0}=0.99$~m. The initial temperatures of the ionized and neutral fluids are set to be uniform at $T_{i0}=T_{n0}=8400$~K to keep the ionization degree the same as that around the solar TMR. 

The initial dimensionless magnetic flux in $z$ direction is given by 
\begin{equation}
  A_{z0}(y)=-b_p\lambda_{\psi} \mathrm{ln} \left[\mathrm{cosh}\left(\frac{y}{\lambda_{\psi}}\right)\right],
\end{equation}
where $b_p$ is the strength of the the magnetic field and $\lambda_{\psi}$ is the initial thickness of the current sheet. The initial magnetic field in z-direction is  
\begin{equation}
  B_{z0}(y)=b_p \bigg/  \left[ \mathrm{cosh}\left(\frac{y}{\lambda_{\psi}}\right) \right].
\end{equation}
In our previous paper \cite{2018ApJ...852...95N}, the numerical results demonstrated that the collisions between electrons and neutrals are not important for magnetic reconnection in low $\beta$ plasmas. In order to compare with the corresponding cases in Ni et al. (2018), we also ignore the collisions between electrons and neutrals in this work. The dimensionless magnetic diffusivity is 
\begin{equation}
   \eta=\eta_{ei}=\eta_{ei\star}T_e^{-1.5},
\end{equation}  
where $\eta_{ei\star}=7.457\times10^{-6}$ is a normalization constant derived from the characteristic values $n_{\star}$, $B_{\star}$, and $L_\star$. $T_e$ is the dimensionless electron temperature. Since the electron and the ion are assumed to be coupled together and only the hydrogen gas is considered in our model, we assume $T_i=T_e$, $n_i=n_e$, and the pressure of the ionized component is twice the ion (or electron) pressure, $P_p = P_e + P_i = 2P_i$. 

The only difference between the model in this work and that in Ni et al. 2018 is the radiative cooling function. In our previous work \cite{2018ApJ...852...95N}, the simple radiative cooling model represents the radiative losses that are due primarily to radiative recombination, with a very crude approximation for radiation due to the presence of excited states of neutral hydrogen. Further, no account is taken of the presence of minority ion species in the TMR.  This simple model is given by
\begin{equation}
  L_{rad} = \Gamma_i^{ion}\phi_{eff},
\end{equation}
where $\phi_{eff}=33$~eV$=5.28\times10^{-18}$~J. The ionization rate $\Gamma_i^{ion}$ is defined as
\begin{equation}
   \Gamma_i^{ion}=  \frac{n_n n_e A}{X+\phi_{\mathrm{ion}}/T_e^{\ast}}\left(\frac{\phi_{\mathrm{ion}}}{T_e^{\ast}}\right)^{K} \mathrm{exp}(-\frac{\phi_{\mathrm{ion}}}{T_e^{\ast}})~~\mbox{m$^{3}$\,s$^{-1}$}, 
\end{equation}
using the values $A=2.91\times10^{-14}$, $K=0.39$, $X=0.232$, and the hydrogen ionization potential $\phi_{\mathrm{ion}}=13.6$~eV. The unit for the neutral and electron number density are both m$^{-3}$, $T_e^{\ast}$ is the electron temperature specified in eV. Then the unit for $\Gamma_i^{ion}$ is m$^{-3}$\,s$^{-1}$ and the unit for $L_{rad}$ is J\,m$^{-3}$\,s$^{-1}$. One finds that the above radiative cooling function approaches zero when a plasma is fully ionized. As shown in our previous work \cite{2018ApJ...852...95N}, the plasmas will be fully ionized if the reconnection magnetic fields are strong enough. This simple radiative cooling model becomes invalid in this situation.

In this work, we use a radiative cooling model computed using the OPACITY project and the CHIANTI databases \citep{2012ApJ...751...75G}. This radiative cooling model is considered to be a more realistic cooling model for plasmas in the solar chromosphere for plasma temperatures below $1.5\times10^4$~K\cite{2017ApJ...842..117A}. The expression for this radiative model in units of J\, m$^{-3}$\,s$^{-1}$ is 
\begin{equation}
  L_{r} = C_E n_e (n_n+n_i) 8.63\times10^{-6}T^{-1/2} \times \sum_{i=1}^2 E_i \Upsilon_i exp(-eE_i/k_BT)~~\mbox{m$^3$\,K$^{1/2}$\,s$^{-1}$}, 
\end{equation}
where $C_E=1.6022\times10^{-25}$~J\,eV$^{-1}$, $E_1=3.54$~eV and $E_2=8.28$~eV, $\Upsilon_1=0.15\times10^{-3}$ and $\Upsilon_2=0.065$. A three level hydrogen atom with two excited levels are included in this function, and $E_1$ and $E_2$ are the excited level energies. In Eq. (6), temperature T is specified in Kelvin and the unit for number density is m$^{-3}$. $k_B=1.3806\times10^{-23}$~J\,K$^{-1}$ is the Boltzmann constant. In the expression for the exponent, $eE_1\simeq 3.54\times1.602\times10^{-19}$~J$\simeq5.671\times10^{-19}$~J, $eE_2\simeq8.28\times1.602\times10^{-19}$~J$\simeq1.326\times10^{-18}$~J. The radiative cooling function for the solar atmosphere with $T\geq2\times10^4$~K can simply take the form \cite{2014masu.book.....P} 
\begin{equation}
  L_{r1} = n_e n_H Q(T), 
\end{equation}
where $Q(T) = 10^{-32} T^{-1/2}$~W\,m$^3$\,K$^{1/2}$ is a reasonable approximation that is useful for analytical modeling over the whole temperature range $2\times10^4$~K$<T<10^7$~K. We have calculated the values of the radiative cooling by using both $L_{r}$ and $L_{r1}$ for $2\times10^4$~K$<T<10^7$~K, the values calculated from the two functions are close for each fixed temperature and plasma density. Therefore, we have used the radiative model $L_r$ provided by Eq. (6) for all the simulations in this work. The background constant heating is also included to balance the initial radiative cooling. The heating function $H_{0}$ is equal to $L_{r0}$ with $n_{e0}$, $n_{n0}$, $n_{i0}$ and $T_{i0}$ set to the initial values shown above. In our simulations, we normalize Eq. (6) by using the characteristic values presented above.      
      
We have simulated three cases in this work, Case~ALr, Case~CLr and Case~ELr. As shown in our previous simulations \cite{2018ApJ...852...95N}, the Hall term and electron-neutral collisions are not important for magnetic reconnection in our model. We have ignored the Hall effect and the electron-neutral collisions in all the three cases in this work. The only difference among Case~ALr, Case~CLr and Case~ELr  is the strength of the initial magnetic field: $b_p=1$ in Case~ALr, $b_p=0.2$ in Case~Clr, $b_p=3$ in Case~Elr. Therefore, one can calculate the initial plasma $\beta$ in each case: $\beta_0=0.058$ in Case~ALr, $\beta_0=1.46$ in Case~CLr, and $\beta_0=0.0064$ in Case~ELr. Except for the radiative function and the background heating, Case~ALr, Case~CLr and Case~ELr  in this work are the same as Case~A, Case~C and Case~E in our previous work \cite{2018ApJ...852...95N}, respectively. The reconnection processes are also symmetric in both x and y direction in Cases~ALr, Case~CLr and Case~ELr. Therefore, we only simulate one quarter of the domain ($0<x<2$, $0<y<1$) in the three cases. We also use the same outer boundary conditions at $|y| = 1$ and the initial electric field perturbations to initiate magnetic reconnection in this work. The perturbation electric field is applied for $0\leq t \leq 1$. The perturbation magnitude is proportional to the value of $b_p$ in each of the cases with the amplitude of $\delta E=10^{-3}b_p$. Periodicity of the physical system is imposed in the x-direction at $|x| = 2$.

The highest number of grid elements in Cases~ALr, CLr, and ELr  is $m_x=96$ elements in the $x$-direction and $m_y=96$ elements in the $y$-direction. We use sixth order basis functions for all simulations, resulting in effective total grid size $(M_x, M_y)=6(m_x,m_y)$. Grid packing is used to concentrate mesh in the reconnection region. Therefore, the mesh packing along the y-direction is concentrated to a thin region near y=0. The quantities shown in the figures in this work are in dimensionless units except for temperatures and velocities.

\section{numerical results}

 In this section, we present the numerical results of simulating magnetic reconnection in the solar TMR with different strengths of initial magnetic field. We compare the results with our previous work \cite{2018ApJ...852...95N} and show how the more realistic radiative cooling model affects the results. The important variables in this work and in our previous paper are listed in table 1. A more in-depth discussion of magnetic reconnection in initially weakly ionized plasmas with different plasma $\beta$ is also presented.  
     
Fig.~1 shows the current density $J_z$ and ionization fraction $f_i$ in one quarter of the domain at three different times in Case~ALr, Case~CLr and Case~ELr respectively. The current sheet lengths in Case~ALr at $t=6.897$, in Case~CLr at $t=21.948$ and in Case~ELr at $t=5.032$ are the same, as are those shown in Fig.~1(b), Fig.~1(e) and Fig.~1(h), and those shown in Fig.~1(c), Fig.~1(f) and Fig.~1(i). As expected, the ionization fraction strongly increases with time inside the current sheet for the cases  (Case~ALr and Case~ELr) with low $\beta$ and strong magnetic field, the ionization fraction slowly increases with time in the high $\beta$ case (Case~CLr). The highest ionization fraction is $72\%$ in Case~ELr, $12\%$ in Case~ALr, and only $0.8\%$ in Case~CLr. Therefore, the lower plasma $\beta$ and higher reconnection magnetic field lead to higher ionization fractions inside the current sheet. However, the ionization fractions in Case~ALr, Case~CLr and Case~ELr are respectively much lower than those in Case~A, Case~C and Case~E in our previous work \cite{2018ApJ...852...95N}. The stronger radiative cooling in this work results in the lower ionization fraction. The neutral fluids in Case~ELr do not become fully ionized as in Case~E with the same plasma $\beta$ in our previous work. 

Fig.~2 shows the profiles of ion and neutral temperatures across the current sheet in Cases~ALr, CLr  and ELr at the same three pairs of times as in Fig.~1. The maximum temperatures within the current sheets in Case~ALr, Case~CLr and Case~ELr are $1.95\times10^4$~K, $1.6\times10^4$~K and $2.3\times10^4$~K, respectively. Therefore, the stronger reconnection magnetic fields and lower plasma $\beta$ result in the higher maximum temperature inside the current sheet. The significant difference of the plasma temperatures between this work and the previous work is the maximum temperature in Case~E and Case~ELr. In the previous work \cite{2018ApJ...852...95N}, the maximum temperature within the narrow current sheet was heated above $4\times10^4$~K in Case~E after the neutral fluids were fully ionized and the simple radiative cooling function was turned off. However, the neutral fluids are not fully ionized in Case~ELr during the reconnection process in this work. Moreover, the strong radiative cooling always exists in Case~ELr even if the plasmas are fully ionized. Therefore, the maximum temperature does not reach above $4\times10^4$~K in Case~ELr. In this work, the ion and neutral temperatures are also nearly equal throughout the evolution due to rapid thermal exchange between the plasma components in all the three cases.   

 In the previous work \cite{2012ApJ...760..109L, 2013PhPl...20f1202L, 2015ApJ...805..134M}, the high $\beta$ simulations showed that the neutral and ionized fluid components decouple upstream of the reconnection site on scales smaller than the neutral-ion mean free path $\lambda_{ni}$. As shown in Fig.~3, the decoupling of neutral and ionized fluid is most obvious in Case~CLr, but the neutral and ion inflows are well coupled in the reconnection phase in Case~ELr. Fig.~3(b) shows that the decoupling of neutral and ion inflows also appears during the later reconnection stage in Case~ALr, which is different from our previous result in Case~A with the same plasma $\beta$ \citep{2018ApJ...852...95N}, the neutral and ion inflows are coupled better in Case~A. The reason for causing such a difference is that the ionized fluid components in Case~ALr are much fewer than that in Case~A. The more ionized components result in a shorter neutral-ion mean free path. Thus, the decoupling of neutral and ion inflows is more obvious in the more weakly ionized plasmas with a longer neutral-ion mean free path. In Fig.~3(a), (d) and (g), one can also see that the ion inflow $V_{iy}$ is higher for a lower $\beta$ case. 

Our simulations results also show that the ionized and neutral fluids are well coupled in the reconnection outflow regions, which is consistent with the previous results \cite{2012ApJ...760..109L, 2013PhPl...20f1202L, 2015ApJ...805..134M, 2018ApJ...852...95N}.  Panels (c), (f) and (i) of Fig.~3 show the outflow plasma velocity $V_{ix}$ at three different times in Case~ALr, CLr and Case~ELr, respectively. The outflow velocities increase with time during the magnetic reconnection process, and are higher in the lower $\beta$ case. The maximum reconnection outflow velocity in Case~ELr is above $50$~km\,s$^{-1}$.

%, which is more than three times higher than that we found in Case~E in our previous work as presented in table 1. However, one should note that the      

Fig.~4(a), (b) and (c) show the four time dependent components contributing to $\partial n_i/\partial t$ in Cases~ALr, CLr and ELr. As in our previous work \cite{2018ApJ...852...95N}, the values of the four corresponding components are the average values inside the current sheet domain at each time.  The ionization rate $\Gamma_i^{ion}$ and the inflow $-\partial (n_i V_{iy})/{\partial y}$ contribute to the gain of ions, and the recombination rate $-\Gamma_i^{rec}$ and the outflow $\partial (n_iV_{ix})/\partial x$ contribute to the loss of ions. These four components behave similarly in Case~ALr, CLr, ELr and Case~A in our previous work \citep{2018ApJ...852...95N}. However, the lower plasma $\beta$ and weaker radiative cooling make the four terms relatively higher. The ionization rate $\Gamma_i^{ion}$ is also faster than the recombination rate $-\Gamma_i^{rec}$ in all of the three cases, which is the same as our previous work \cite{2018ApJ...852...95N} but  significant different from the previous higher $\beta$ simulations \cite{2012ApJ...760..109L, 2013PhPl...20f1202L}. We have also tested a simulation with the initial magnetic field four times smaller than that in Case~CLr; the ionization rate is eventually smaller than the recombination rate in such a high $\beta$ case ($\beta_0=23.36$). Therefore, the plasma $\beta$ inside the current sheet region appears to be the main factor determining whether ionization or recombination dominate within the current sheet during the reconnection process. From the simulation results presented in this work, we conclude that the ionization rate $\Gamma_i^{ion}$ becomes faster than the recombination rate $-\Gamma_i^{rec}$ somewhere between $\beta=1.46$ and $\beta=23.36$ inside the current sheet.  

The initial radiative cooling $L_r$ in Eq. (6) in this work and $L_{rad}=\Gamma_i^{ion} \phi_{eff}$ in our previous work \cite{2018ApJ...852...95N}  have both been calculated. Their respective dimensionless values are $L_{r0}=5.6927\times10^{-6}$ and $L_{rad0}=6.7225\times10^{-9}$. Therefore, the initial radiative cooling in this work is about three orders of magnitude higher than that in our previous work \cite{2018ApJ...852...95N}. One should notice that the background heating is also included in this work, but the radiative cooling inside the current sheet increases sharply with time and quickly becomes much greater than the background heating. In this work, we have calculated the time evolution of the total radiated energy $Q_{rad}=\int_{0}^{1}\int_{0}^{2} L_{rad} dxdy$, the total background heating $Q_{bh}=\int_{0}^{1}\int_{0}^{2} H_{0} dxdy$, the Joule heating $Q_{Joule}=\int_{0}^{1}\int_{0}^{2} \eta J^2 dxdy$, the frictional heating between ions and neutral particles $Q_{in}$ and the viscous heating of ions and neutral particles $Q_{vis}$ in the whole simulation domain. Joule heating is several orders of magnitude higher than other heating terms in all of our simulations (not shown). Fig.~5 shows the time evolution of $Q_{Joule}+Q_{in}+Q_{vis}$, $Q_{rad}$ and $Q_{bh}$ in Case~ALr, CLr and ELr. It is shown that most of the total generated thermal energy $Q_{Joule}+Q_{in}+Q_{vis}$ is radiated in all of the three cases. The stronger reconnection magnetic fields result in the more generated Joule heating, the radiative cooling is also becoming much greater when more neutrals inside the current sheets are ionized. Therefore, both the Joule heating and radiative cooling in Case~ELr are the greatest in all of the three cases. Comparing Case~ALr and CLr with Case~A and C in our previous work \cite{2018ApJ...852...95N}, one can find that both the generated thermal energy and radiated heat in Case~ALr are correspondingly a little bit higher than those in Case~A at the same time point, the radiated heat in Case~CLr is obviously higher than that in Case~C. The stronger radiative cooling model in this work makes the neutrals inside the current sheet more difficult to be ionized for the same reconnection magnetic fields and plasma $\beta$. Though the initial radiative cooling in this work is about three orders of magnitude higher than that in our previous work \cite{2018ApJ...852...95N}, the values of the radiated heat in Case~ALr, CLr and ELr are correspondingly at the same order of magnitude as those in Case~A, C and E during the reconnection process before the plasmas are nearly fully ionized. In the reconnection process, the strong ionization rate $\Gamma_i^{ion}$ in the previous work makes the simple radiative cooling  $L_{rad}=\Gamma_i^{ion} \phi_{eff}$ to be big enough to compare with the radiative cooling applied in this work for the same plasma $\beta$. Fig.~5 also shows that the background heating $Q_{bh}$ is very small compared with $Q_{rad}$ and $Q_{Joule}+Q_{in}+Q_{vis}$, it can be ignored during the magnetic reconnection process. In the paper by Alvarez Laguna et al. 2017, the heating term was the same order as the radiative cooling term, including the background heating strongly affected the reconnection process in their work \cite{2017ApJ...842..117A} .     

Fig.~6(a) shows the half length $L_{sim}$ and width $\delta_{sim}$ of the CSs in Case~ALr, CLr, and ELr. The half length of the current sheet is also about $0.5-0.6$ during the later stages of magnetic reconnection in all of the three cases. The half width of the current sheet eventually drops to about $0.012$ in Case~ALr, $0.017$ in Case~CLr and $0.008$ in Case~ELr. The reconnection rate is calculated as $M_{sim}=\eta^{\ast} j_{max}/(V_A^{\ast} B_{up})$, where $j_{max}$ is the maximum value of the out of plane current density $j_{z}$, located at $(x,y)=(0,0)$ in all the simulations in this work. $B_{up}$ is $B_x$ evaluated at $(0, \delta_{sim})$, where $\delta_{sim}$ is defined as the half-width at half-max in $j_{z}$. $V_{A}^{\ast}$ is the relevant  Alfv\'en velocity defined using $B_{up}$ and the total number density $n^{\ast}$ at the location of $j_{max}$. $\eta^{\ast}$ is the magnetic diffusion coefficient defined in Equation (3) at the location of $j_{max}$. The solid lines in Fig.~6(b) represent the time evolution of the reconnection rates in Cases~ALr, CLr and  ELr.  The reconnection rate $M_{sim}$ can reach 0.121 in Case~CLr, which is the highest in all of the three cases. The maximum reconnection rate in Case~ELr is only around 0.016, which is the lowest in all of the three cases. We have also calculated the reconnection rates $M_{SP}$ which are predicted by the Sweet-Parker model, $M_{SP}=1/\sqrt{S_{sim}}$. The Lundquist number $S_{sim}$ in the simulations is defined by $S_{sim}=V_{A}^{\ast}L_{sim}/\eta^{\ast}$. The dash-dotted lines in Fig.~6(b) represent $M_{SP}$ in the three cases. One can find that the value of $M_{SP}$ predicted by the Sweet-Parker model is about three times smaller that the realistic reconnection rate $M_{sim}$ during the later reconnection stage in Case~CLr. The values of the reconnection rate $M_{sim}$ are much closer to the values of $M_{SP}$ in Case~ALr and ELr. As discussed above and shown in Fig.~3, the decoupling of ion and neutral inflows is obvious only in Case~CLr with a much higher plasma $\beta$. Therefore, the decoupling of ion and neutral inflows can result in a much faster reconnection process than that predicted by the Sweet-Parker model in Case~CLr. The Sweet-Parker type magnetic reconnection appears in Case~ALr and ELr. 

Alvarez Laguna et al. (2017) concluded that strong radiative cooling produced faster reconnection than without radiation \cite{2017ApJ...842..117A} . However, the strong radiative cooling in Case~ALr and ELr in our simulations dose not result in a faster reconnection than that predicted by the Sweet-Parker model. One should notice that the recombination dominated over the ionization in all the high $\beta$ simulations \cite{2017ApJ...842..117A} in Alvarez Laguna et al. (2017). The strong radiative cooling resulted that the ionization degrees at the reconnection X-points sharply decreased with time in the cases with high initial ionization degrees, the recombination effect and the decoupling of ion and neutral inflows were even more significant in the cases with strong radiative cooling than those without radiative cooling. Though the same strong radiative cooling model is included in our simulations,  the ionization degrees at the reconnection X-points increase with time and the ionization is always faster than recombination in Cases~ALr, CLr and ELr in our simulations, especially in Case~ALr and ELr with very low plasma $\beta$ and strong reconnection magnetic fields. The decoupling of ion and neutral inflows only significantly appears in Case~CLr. We can conclude that the decoupling of ion and neutral inflows is not significant and the recombination does not obviously affect magnetic reconnection in strongly magnetized regions (above 500 G) around the solar TMR.  

As presented in Tabel~1, the maximum current densities and reconnection outflow velocities in Case~ALr and ELr are higher than those in Case~A and E, respectively. However, these variables increase with time in our simulations. The runs in Case~ALr and  ELr lasted longer than those in Case~A and E. Therefore, the maximum current densities and reconnection outflow velocities could have reached higher values in Cases~A and E if the simulation runs lasted longer. The current sheet width in Case~E reached a very small value ($0.003L_0$) and the maximum reconnection rate was higher than that in Case~A.  However, one should note that the radiative cooling model in the prior work became particularly unrealistic in situations such as that in the latter stages of the Case~E simulation, when the hydrogen plasma in the current sheet became nearly fully ionized.

\begin{table}
 \caption{The important variables in Case~ALr, CLr and ELr in this work and in Case~A, C, E in our previous work \cite{2018ApJ...852...95N}.The maximum values of the ionization fraction $f_i$, the plasma temperature $T_i$, the current density $J_z$, the difference between the ion and neutral inflows $\vert V_{iy}-V_{ny} \vert$, the outflow ion velocity $V_{ix}$, the magnetic reconnection rate $M_{sim}$,  the minimum value of the current sheet width $\delta_{sim}$ and the ion density $n_i$ during the later reconnection stage are presented.} 
 \label{models}  
 \begin{tabular}{|c|c|c|c|c|c|c|c|cl} 
 \hline
                              & Max ($f_i$) & Max ($T_i$) & Max ($J_z$) & Max ($\vert V_{iy}-V_{ny} \vert$)  & Max ($V_{ix}$) & Max ($M_{sim}$) & $\delta_{sim}$ & Max($n_i$)  \\  
 \hline \hline 
                       Case ALr   & $12\%$ &  $1.95\times10^4$~K& 44  &        0.185 km/s   & 21.1 km/s   &       0.025              &  0.012   & 0.0860                       \\  \cline{1-9} 
                        Case CLr   & $ 0.8\%$&  $1.6\times10^4$~K &  6    &       0. 692 km/s  &  6.3 km/s &      0.121                  &   0.017   & 0.0027                  \\   \cline{1-9} 
                       Case ELr    & $  72\% $& $2.3\times10^4$~K & 206 &      0.076 km/s     &  52.3 km/s  &       0.016             &   0.008    & 0.5884                  \\  \cline{1-9} 
                  
 \hline \hline
                       Case A   & 45\%& $1.6\times10^4$~K &  29     &    0.048 km/s        &   13.9 km/s    &        0.030          &       0.015   & 0.2873                 \\   \cline{1-9} 
                       Case C  &  3\%   & $1.6\times10^4$~K &   5.5    &  0.182 km/s        &    6.7 km/s &     0.109          &     0.018    &    0.0094            \\   \cline{1-9} 
                      Case E    &  100\%& $4.6\times10^4$~K&  192   &  0.061 km/s        &   16.4 km/s &   0.035          &     0.003     &   0.5133             \\  \cline{1-9} 
   
 \hline 
\end{tabular} 
\end{table}

\begin{figure}
      \centerline{\includegraphics[width=0.33\textwidth, clip=]{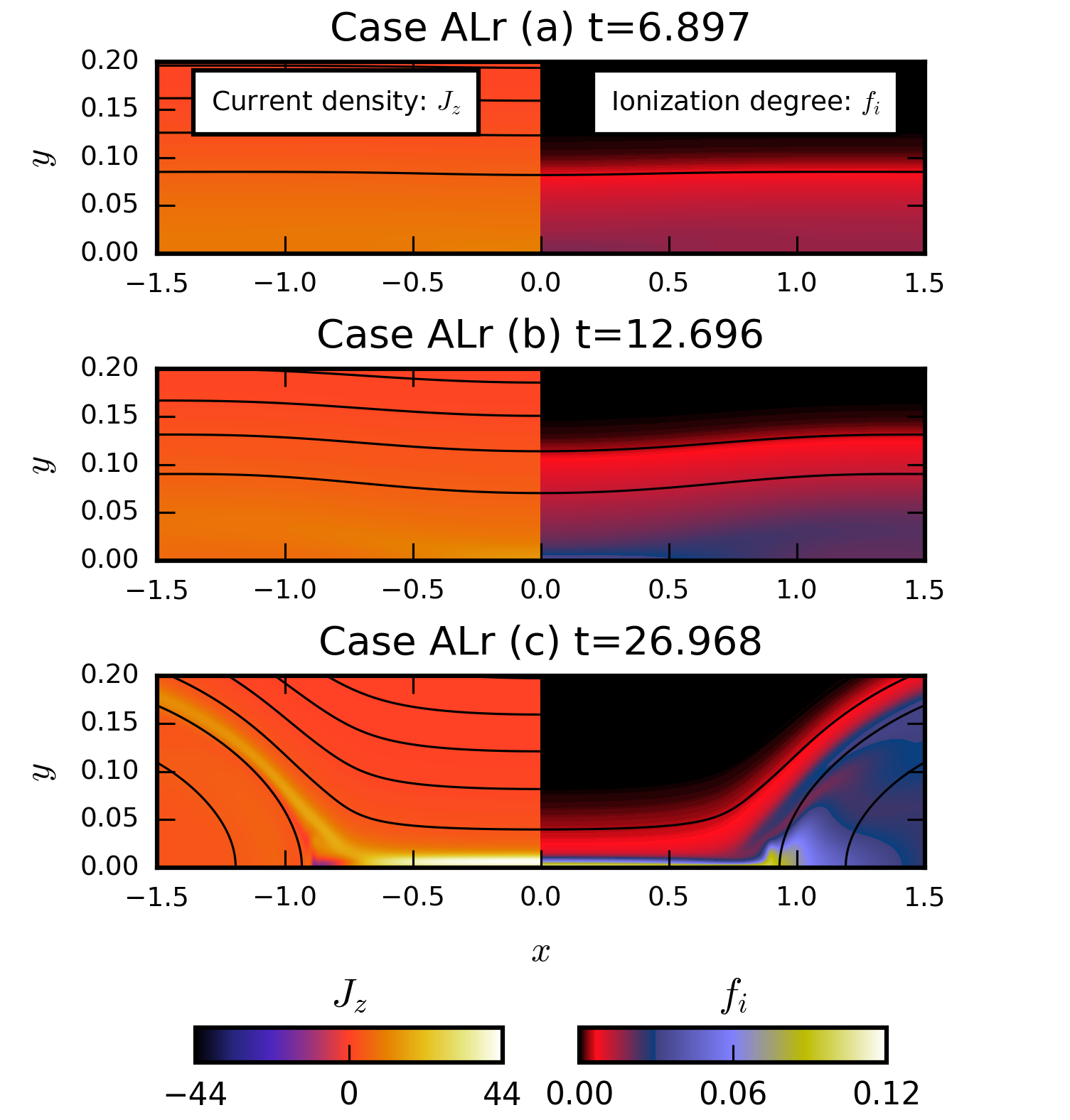}
                          \includegraphics[width=0.33\textwidth, clip=]{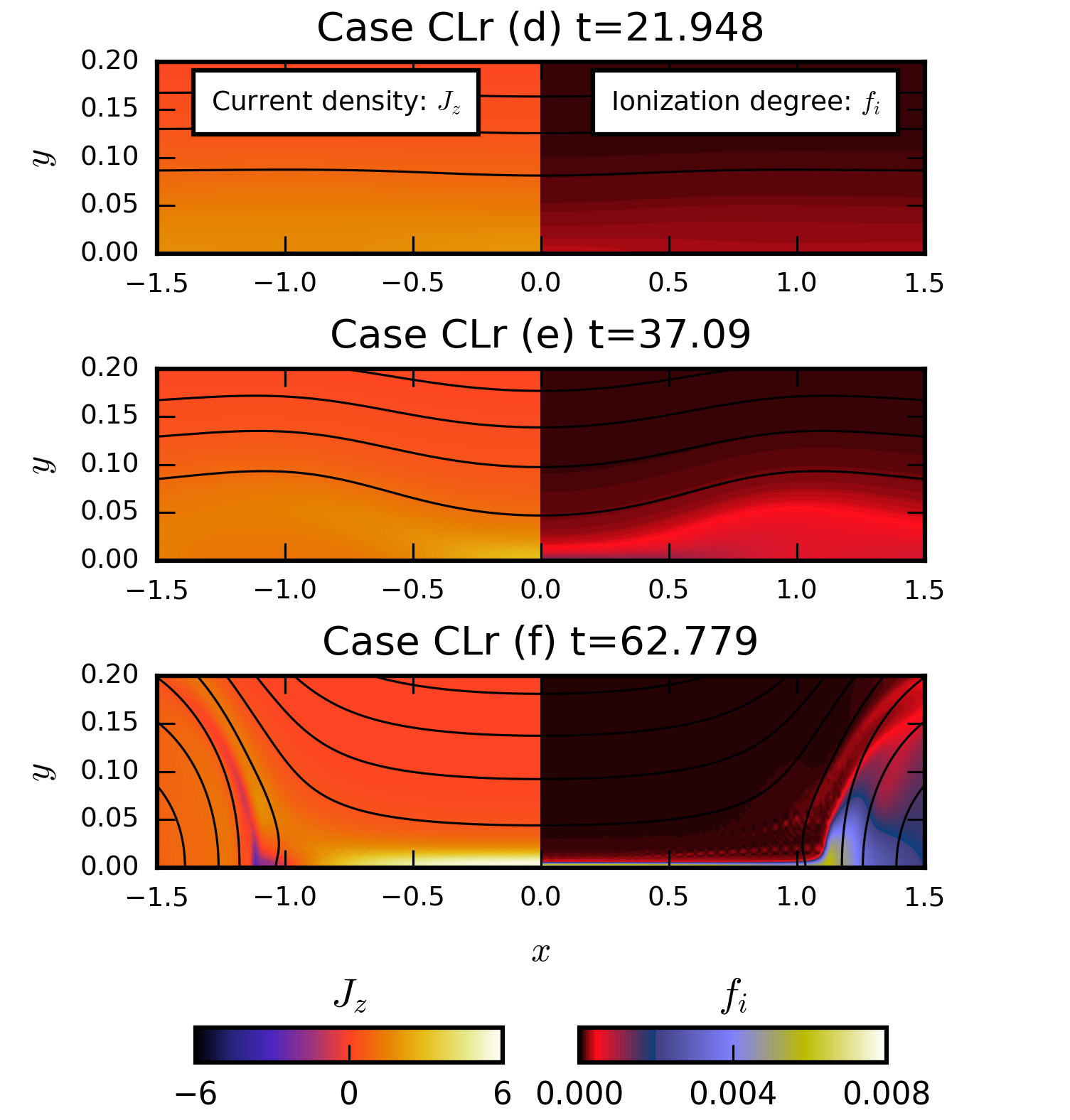}
                          \includegraphics[width=0.33\textwidth, clip=]{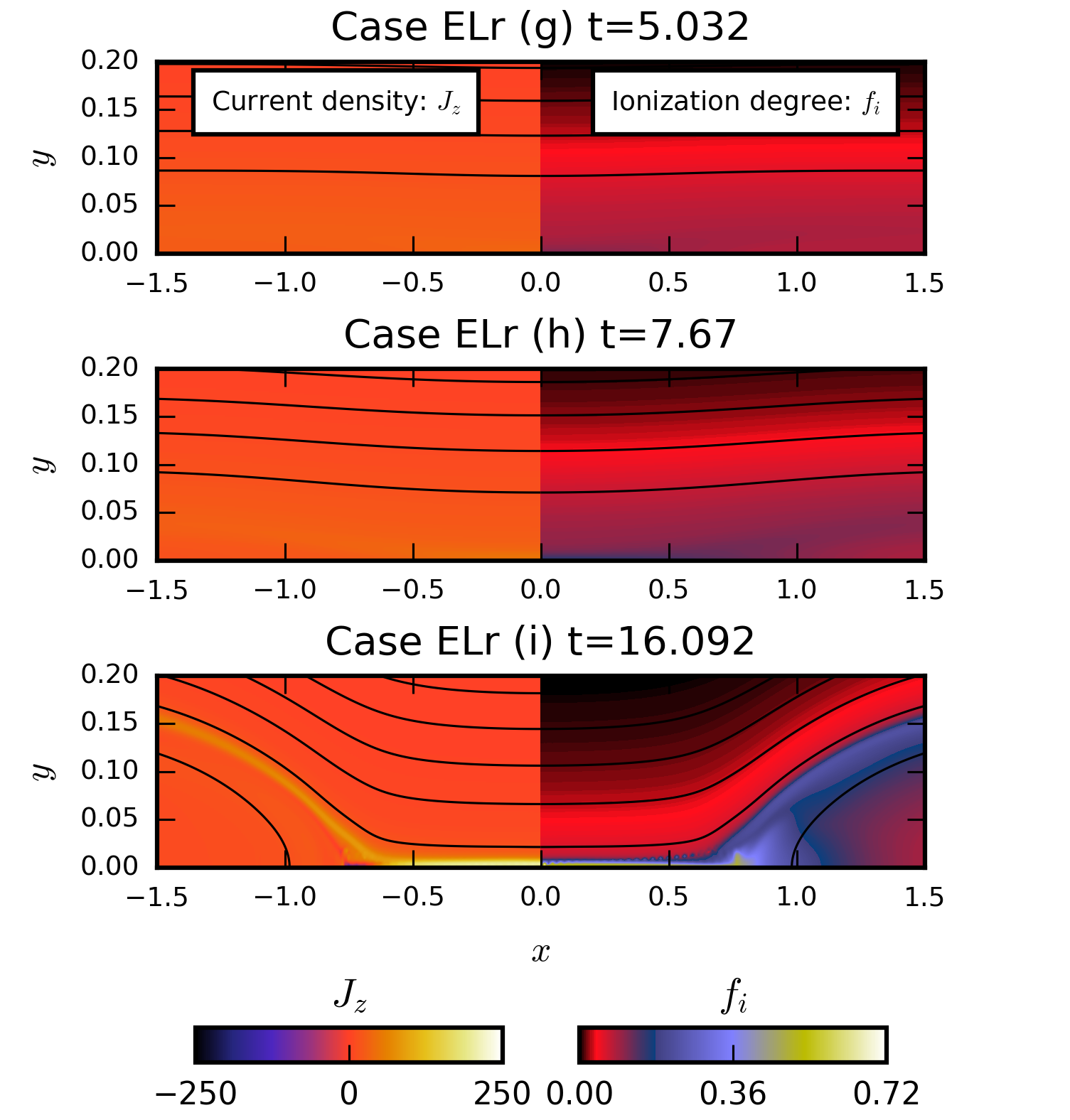}}
      \caption{(a), (b) and (c) show the current density $J_z$ (left) and ionization degree $f_i$ (right)  in one quarter of the domain at $t=6.897$, $t=12.696$ and $t=26.968$ in Case~ALr; (d),(e) and (f) show the same at $t=21.948$, $t=37.09$ and $t=62.779$ in Case~CLr; (g),(h) and (i) show the same at $t=5.032$, $t=7.67$ and $t=16.092$ in Case~ELr. The black contour lines represent the out of plane component of the magnetic flux $A_z$ in these 2D figures. }
    \label{fig.1} 
\end{figure}
 
 \begin{figure}
      \centerline{\includegraphics[width=0.33\textwidth, clip=]{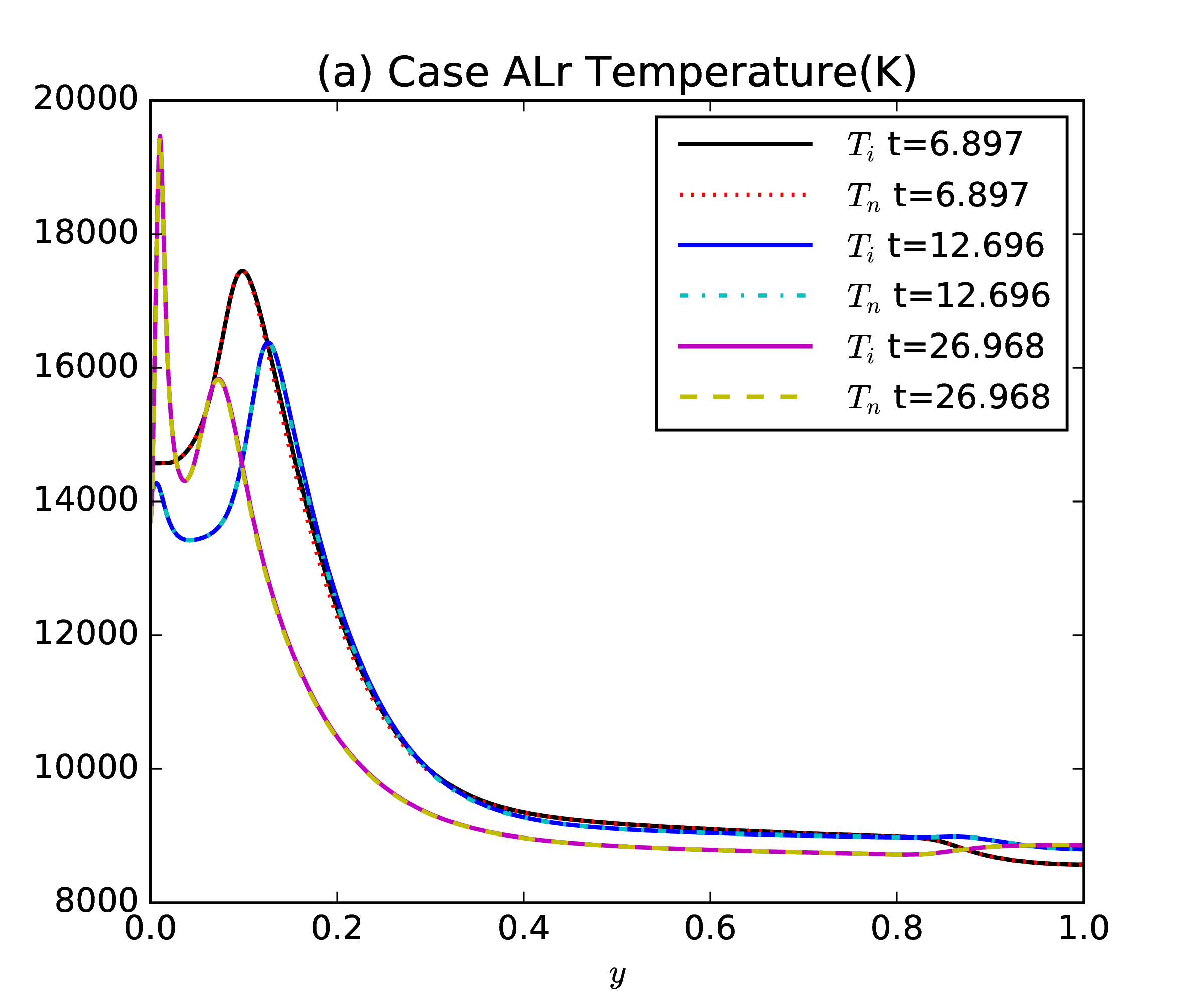}
                          \includegraphics[width=0.33\textwidth, clip=]{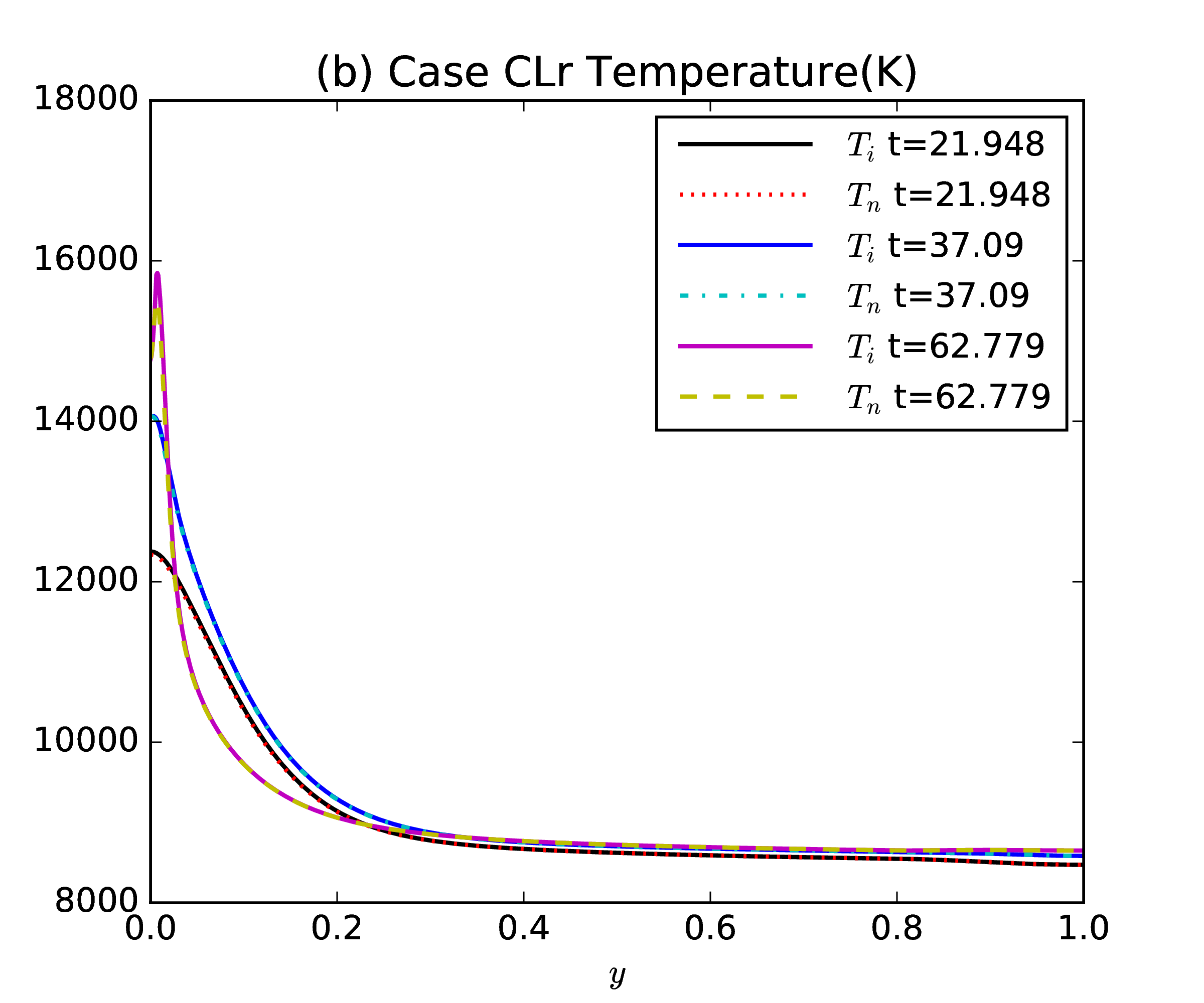}
                          \includegraphics[width=0.33\textwidth, clip=]{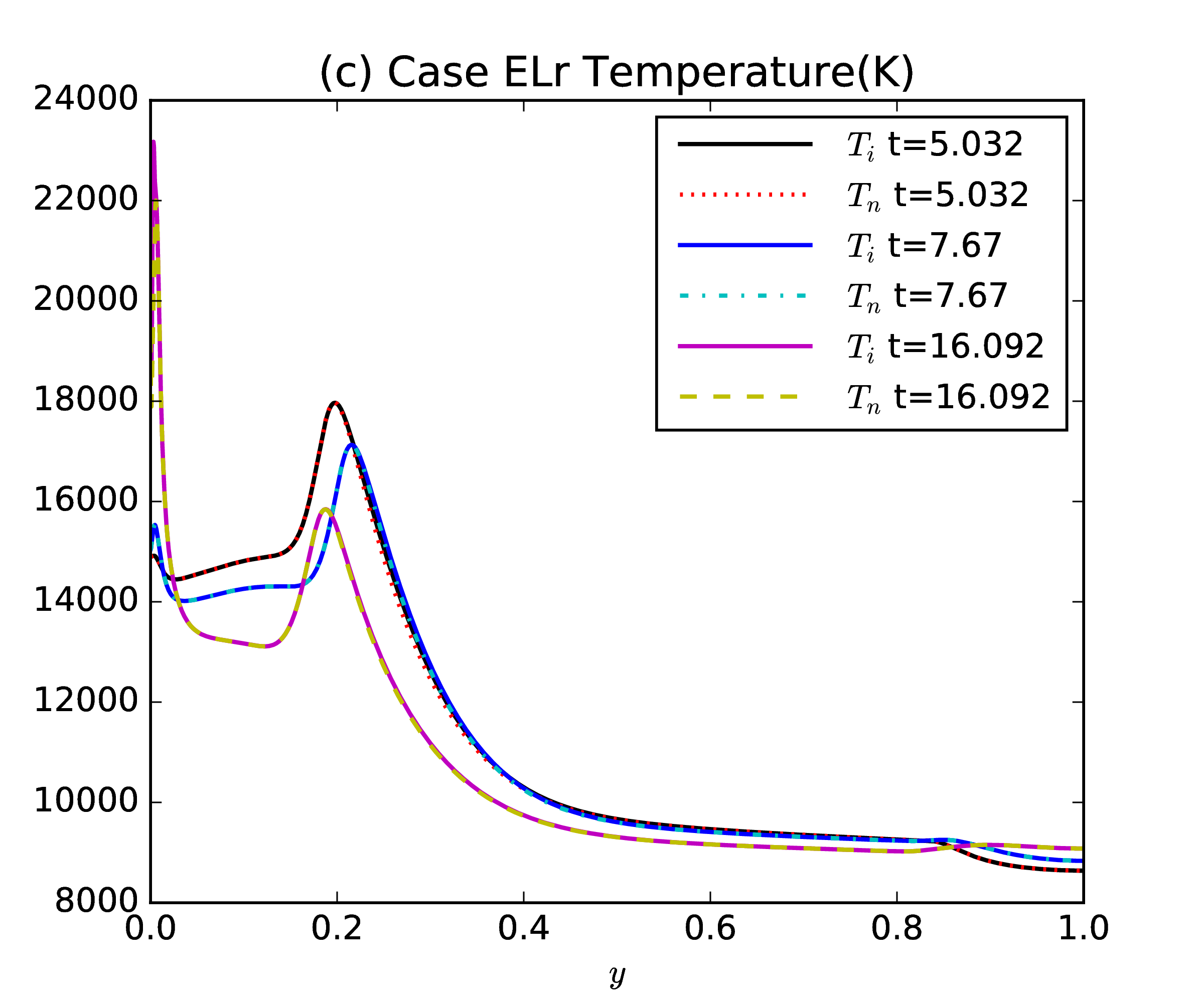} }
      \caption{(a) shows the distributions of the ion temperature $T_i$ and neutral temperature $T_n$ in Kelvin at $x=0$ along y direction at $t=6.897$, $t=12.696$ and $t=26.968$ in Case~ALr; (b) shows the same at $x=0$ along y direction at $t=21.948$, $t=37.09$ and $t=62.779$ in Case~CLr; (c) shows the same at $x=0$ along y direction at $t=5.032$, $t=7.67$ and $t=16.092$ in Case~ELr. }
    \label{fig.2}
 \end{figure} 
   
 \begin{figure}
    \centerline{\includegraphics[width=0.33\textwidth, clip=]{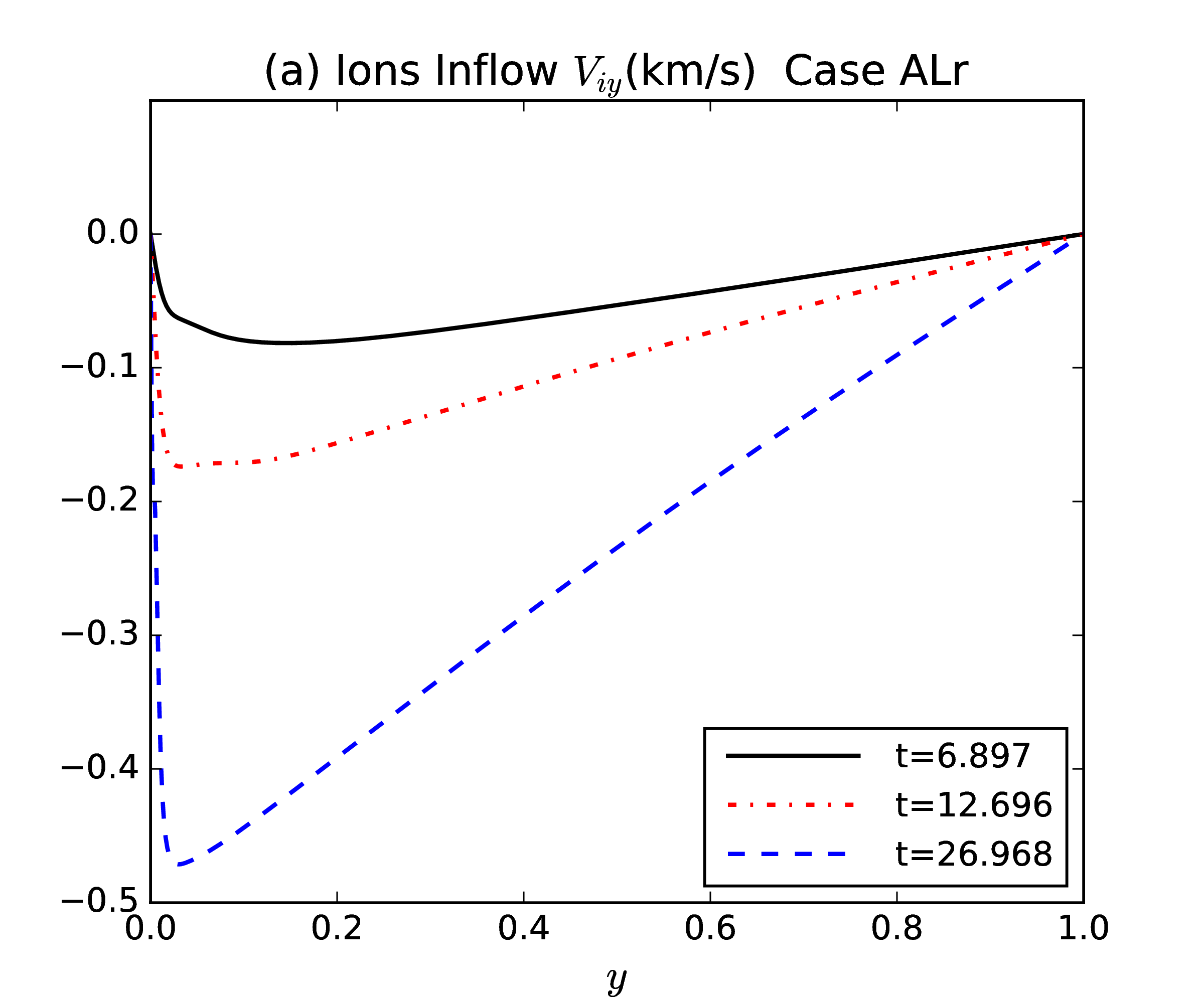}
                         \includegraphics[width=0.33\textwidth, clip=]{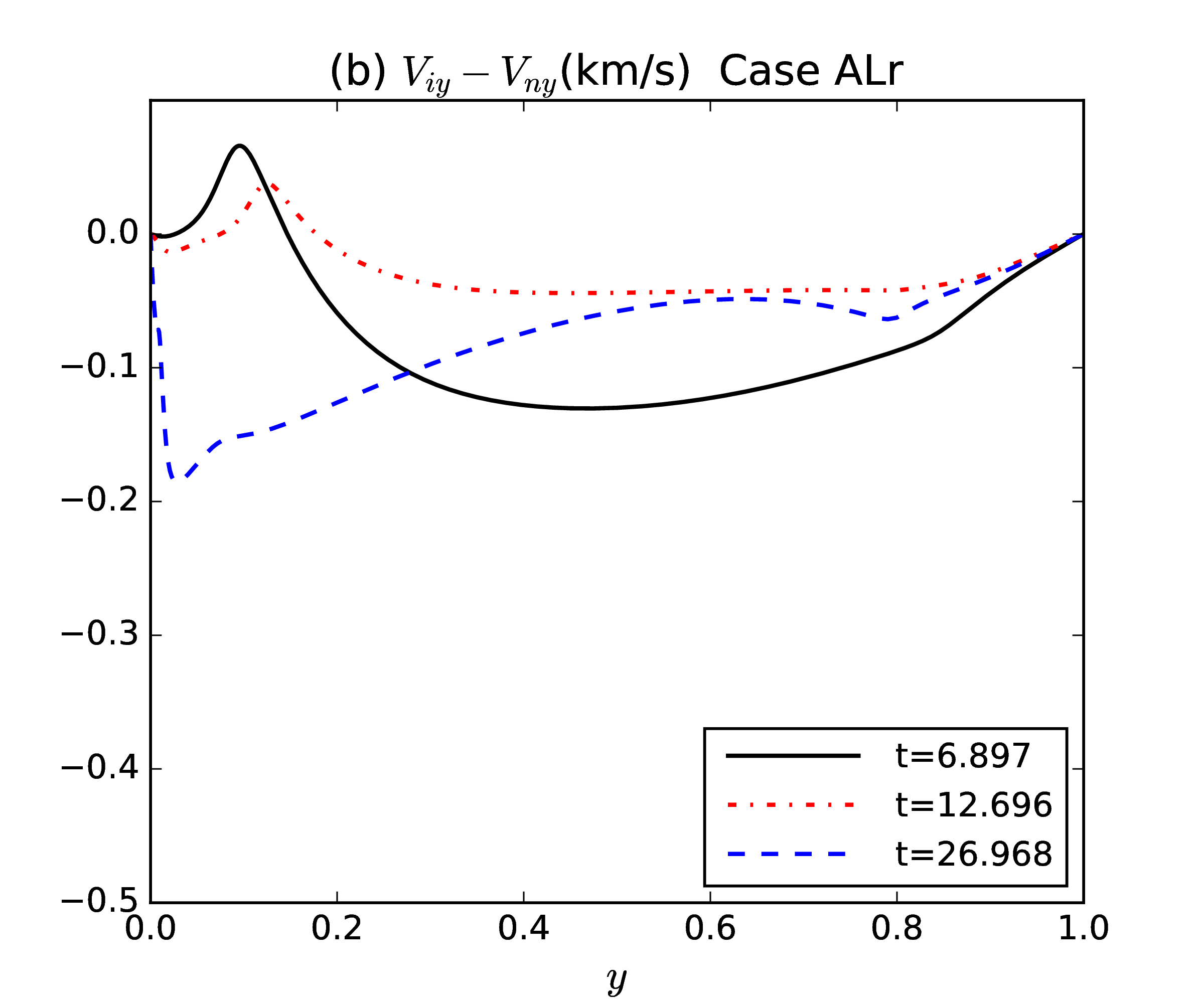} 
                         \includegraphics[width=0.33\textwidth, clip=]{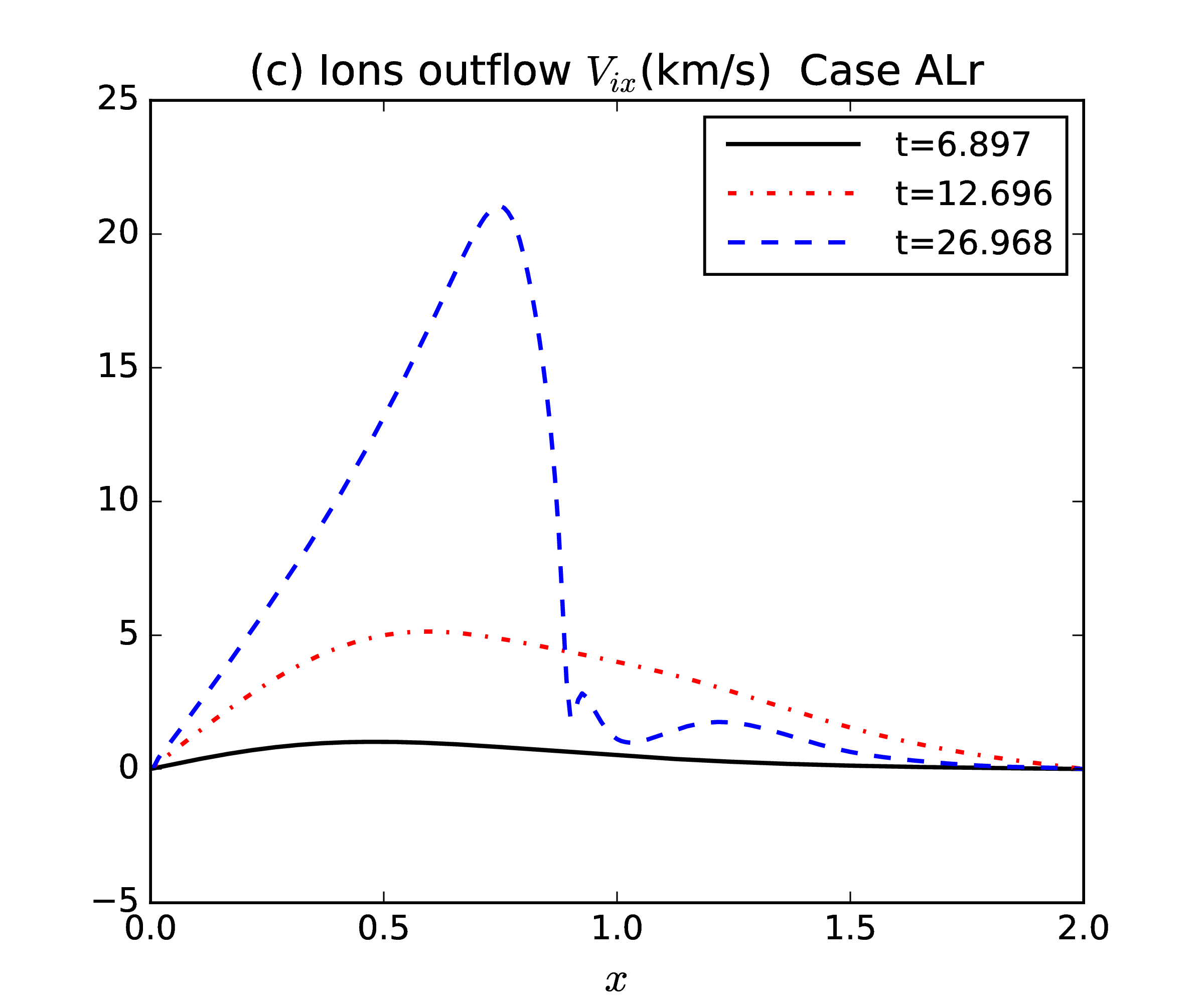}}
   \centerline{\includegraphics[width=0.33\textwidth, clip=]{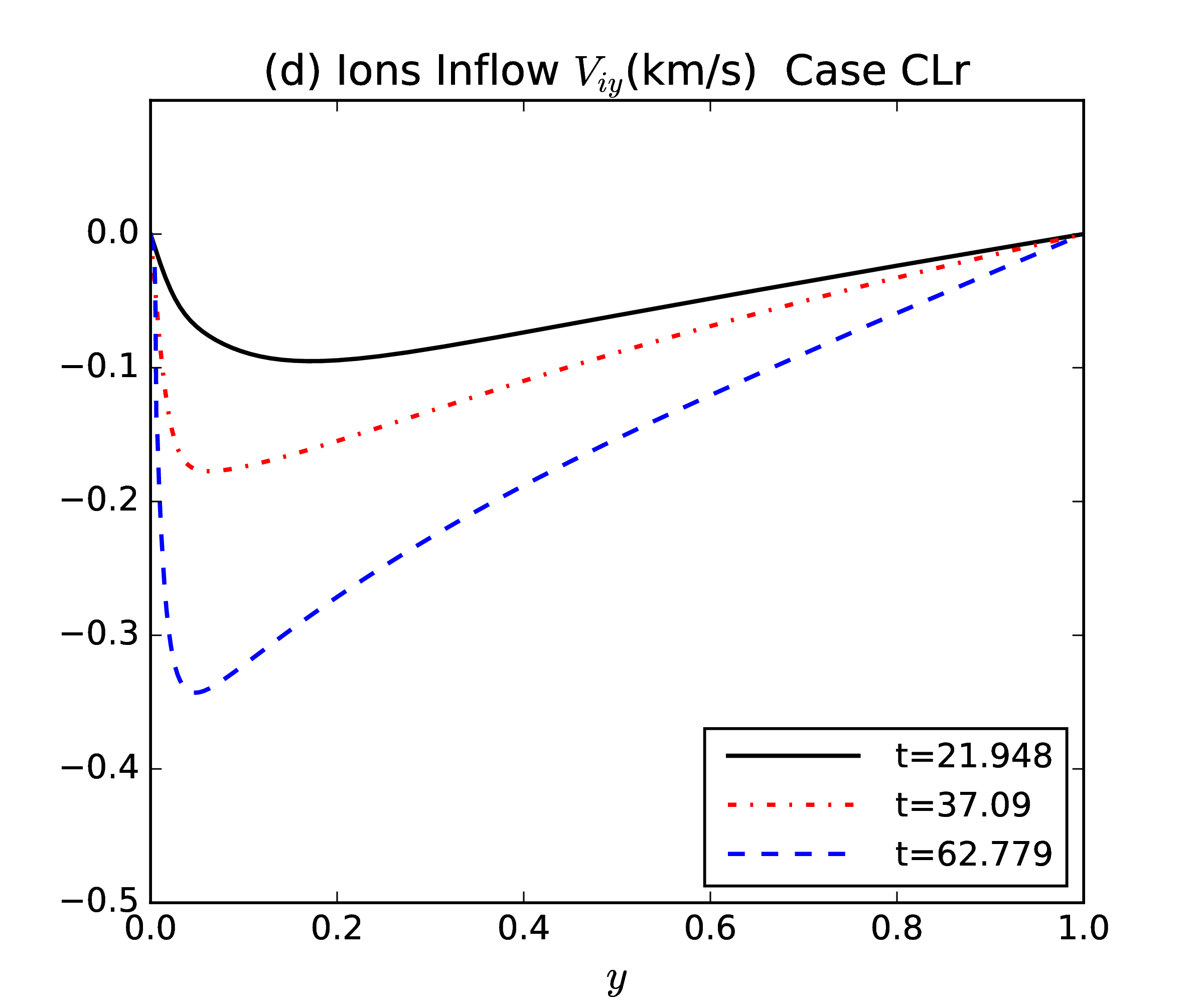}
                      \includegraphics[width=0.33\textwidth, clip=]{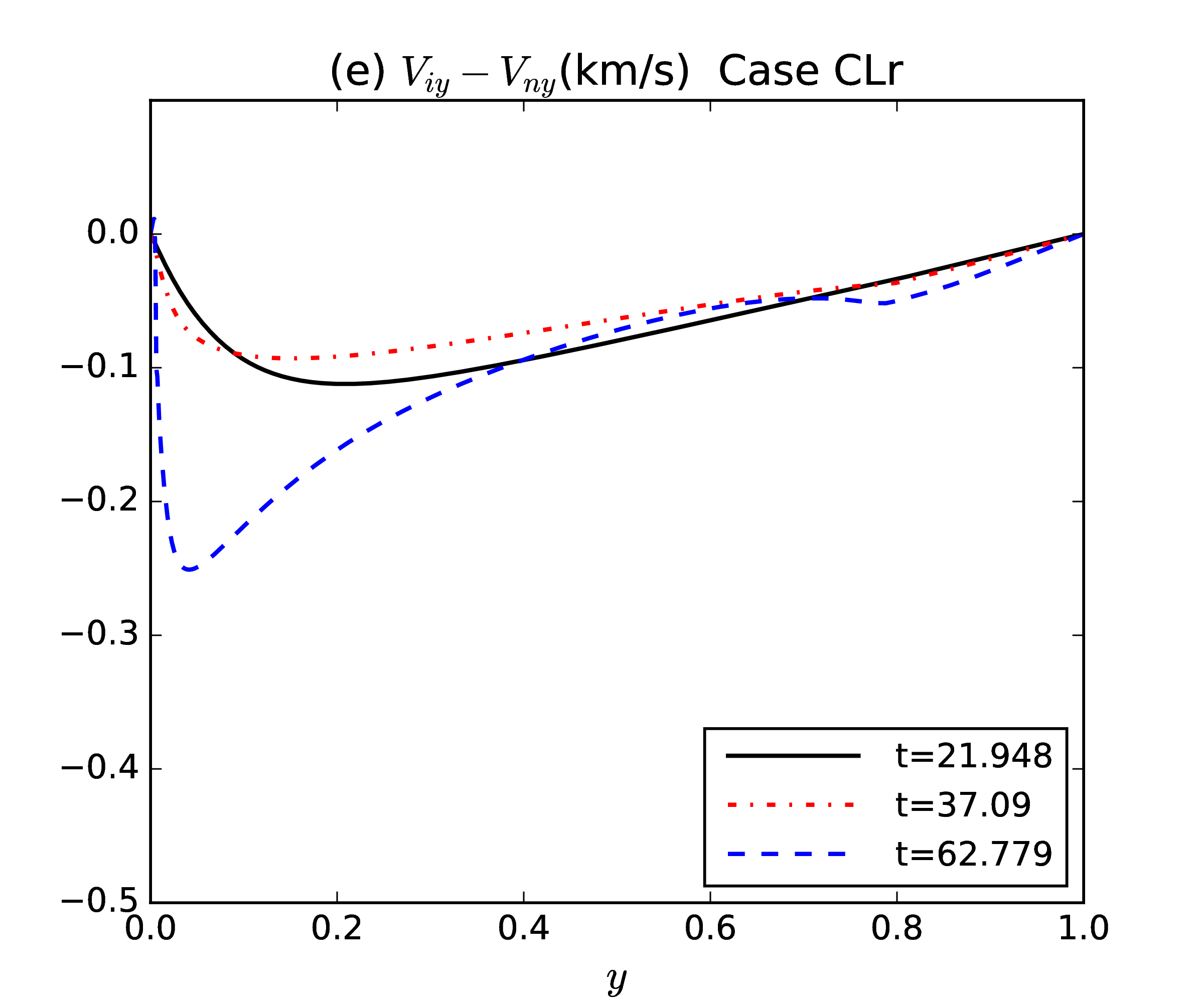} 
                           \includegraphics[width=0.33\textwidth, clip=]{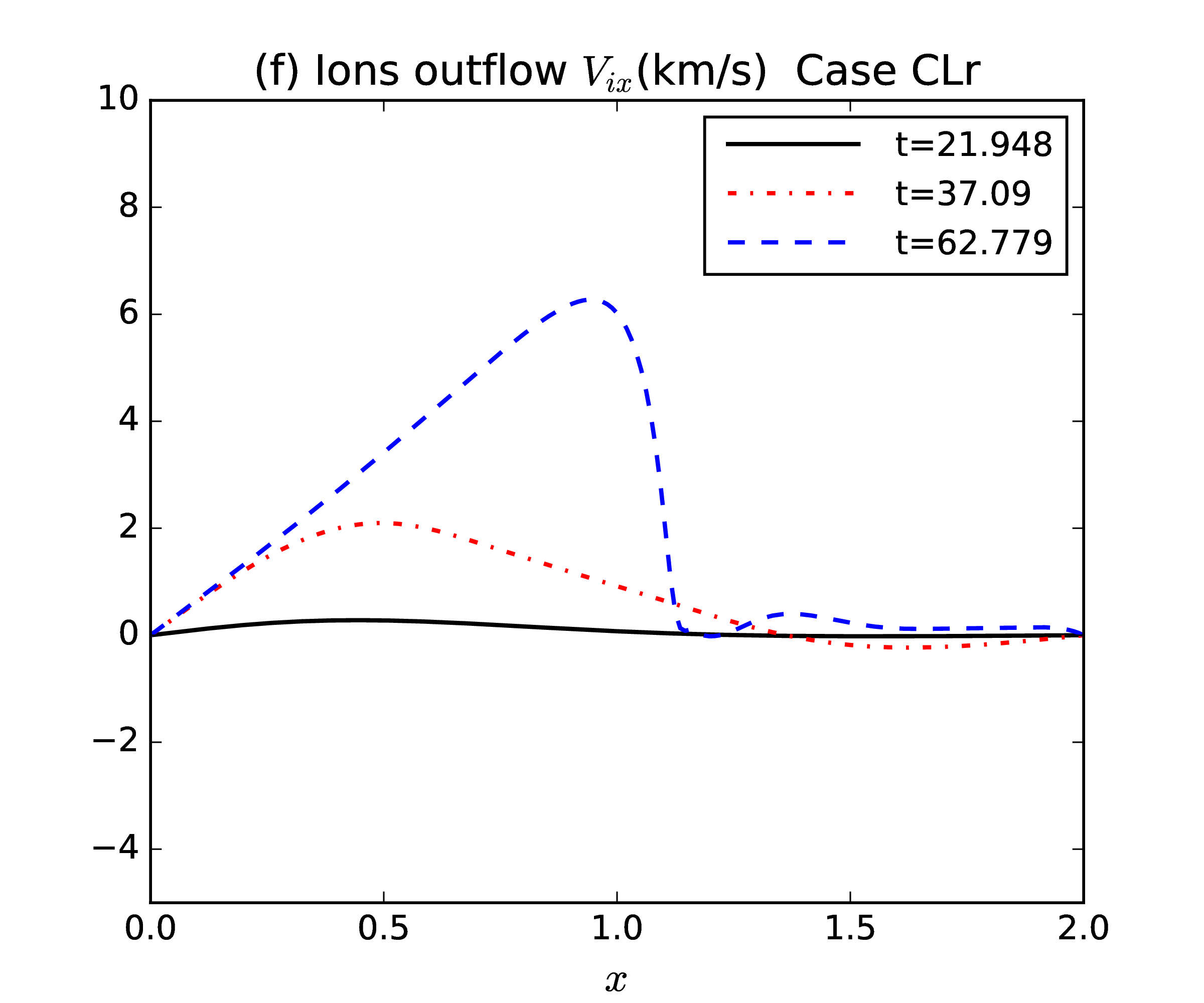}} 
    \centerline{\includegraphics[width=0.33\textwidth, clip=]{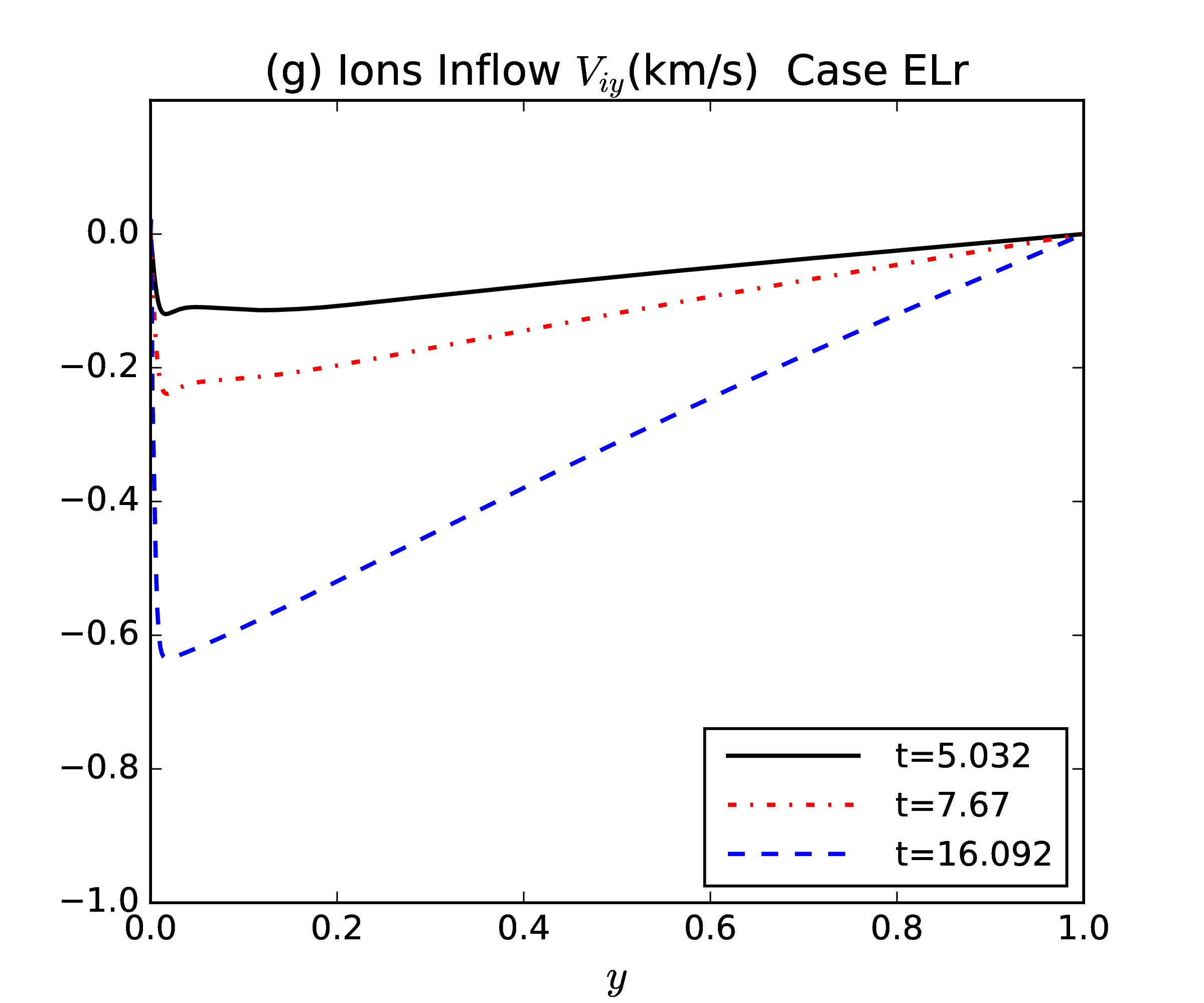}
                      \includegraphics[width=0.33\textwidth, clip=]{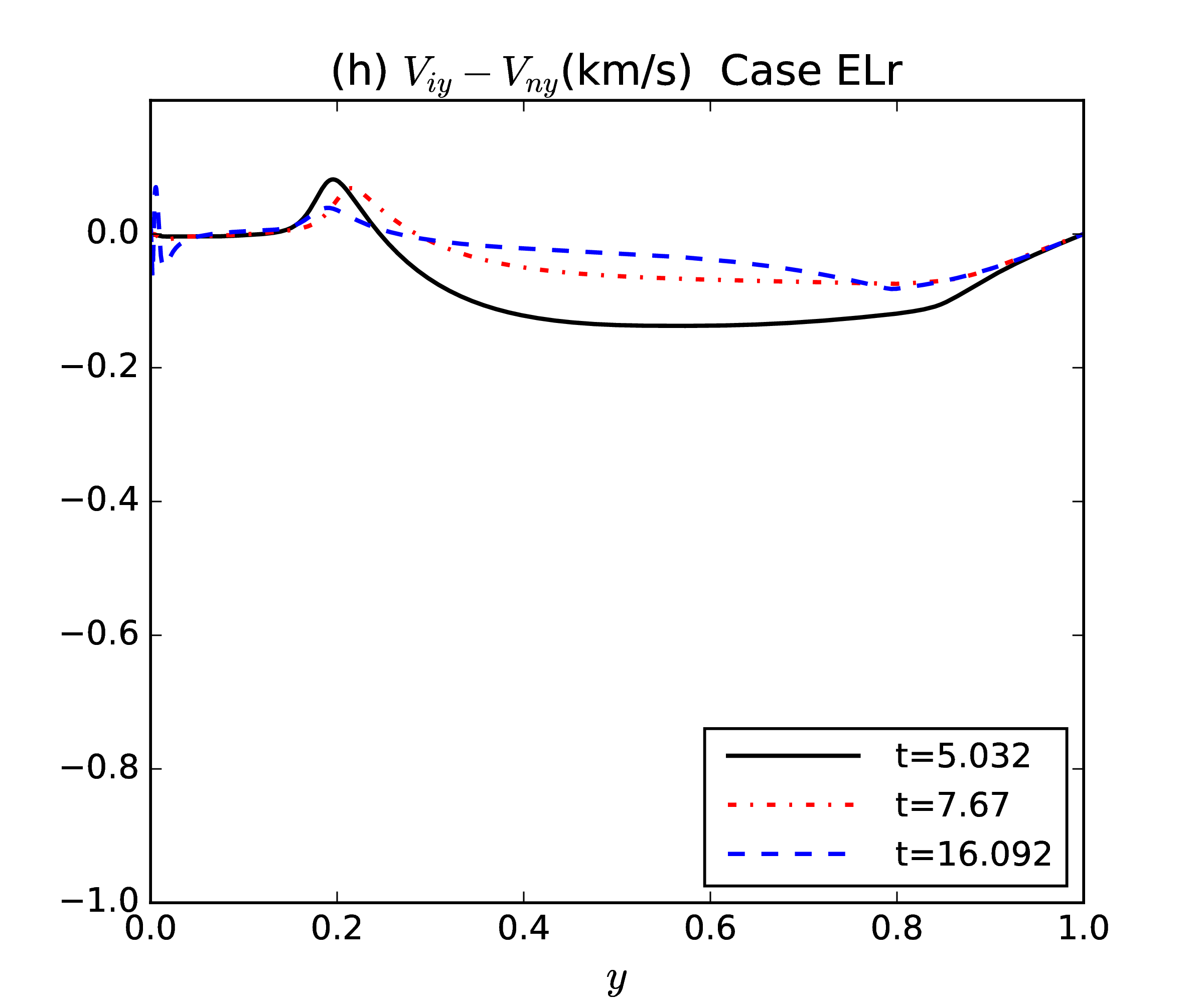} 
                           \includegraphics[width=0.33\textwidth, clip=]{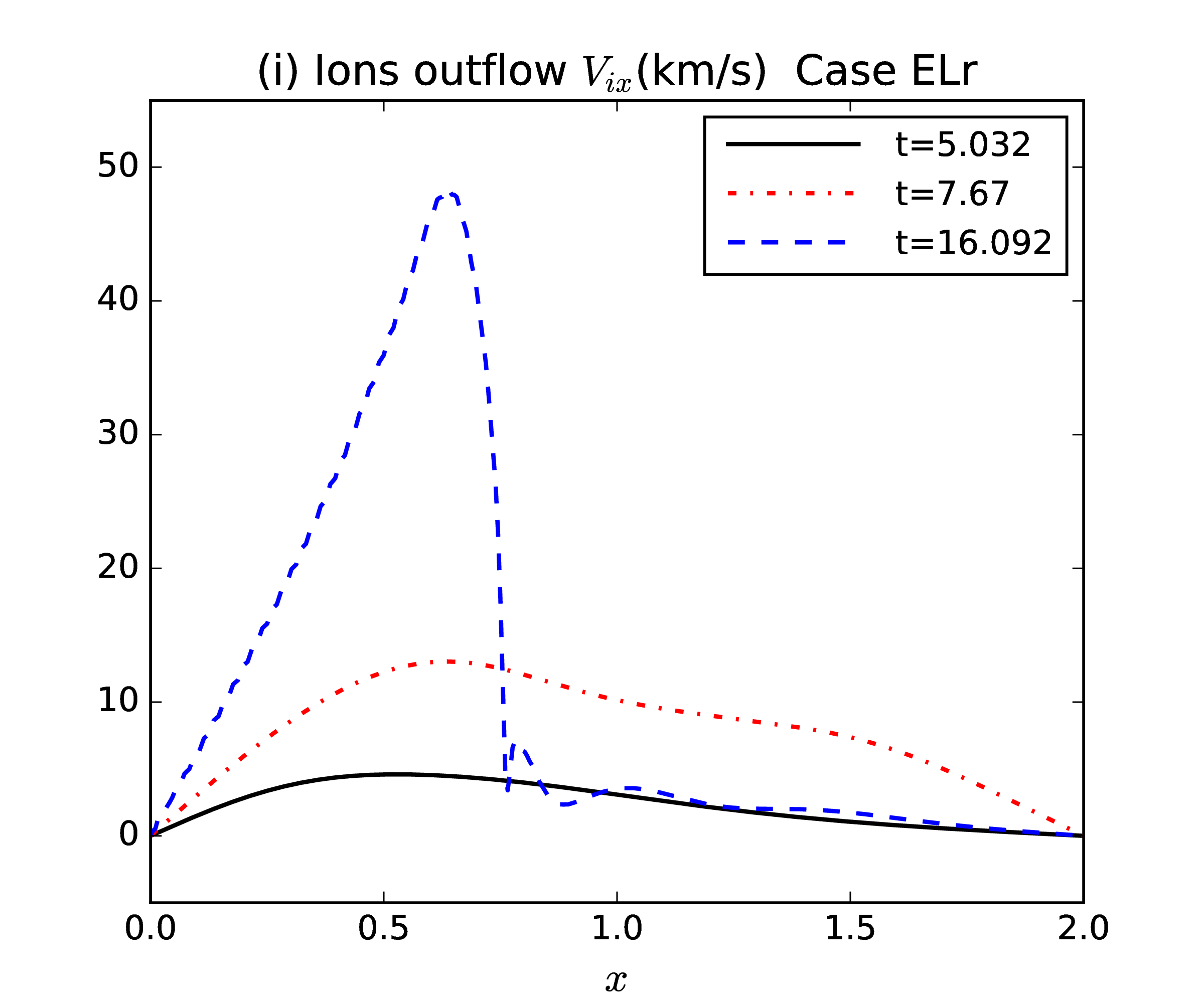}}
     \caption{Panels (a), (d) and (g) show the ion inflow speed $V_{iy}$ cross the current sheet with dimensions  at $x=0$ in Case~ALr, CLr and
     ELr. Panels (b), (e) and (h) show the difference in speed between the ion and the neutral inflows $V_{iy}-V_{ny}$ across the current sheet 
     with dimensions at $x=0$ in Case~ALr, CLr and ELr. Panels (c), (f) and (i) show the ion outflow $V_{ix}$ along the current sheet at  
     $y=0$ in Case~ALr, CLr and ELr.}
     \label{fig.3}
\end{figure} 
  
  \begin{figure}
      \centerline{\includegraphics[width=0.33\textwidth, clip=]{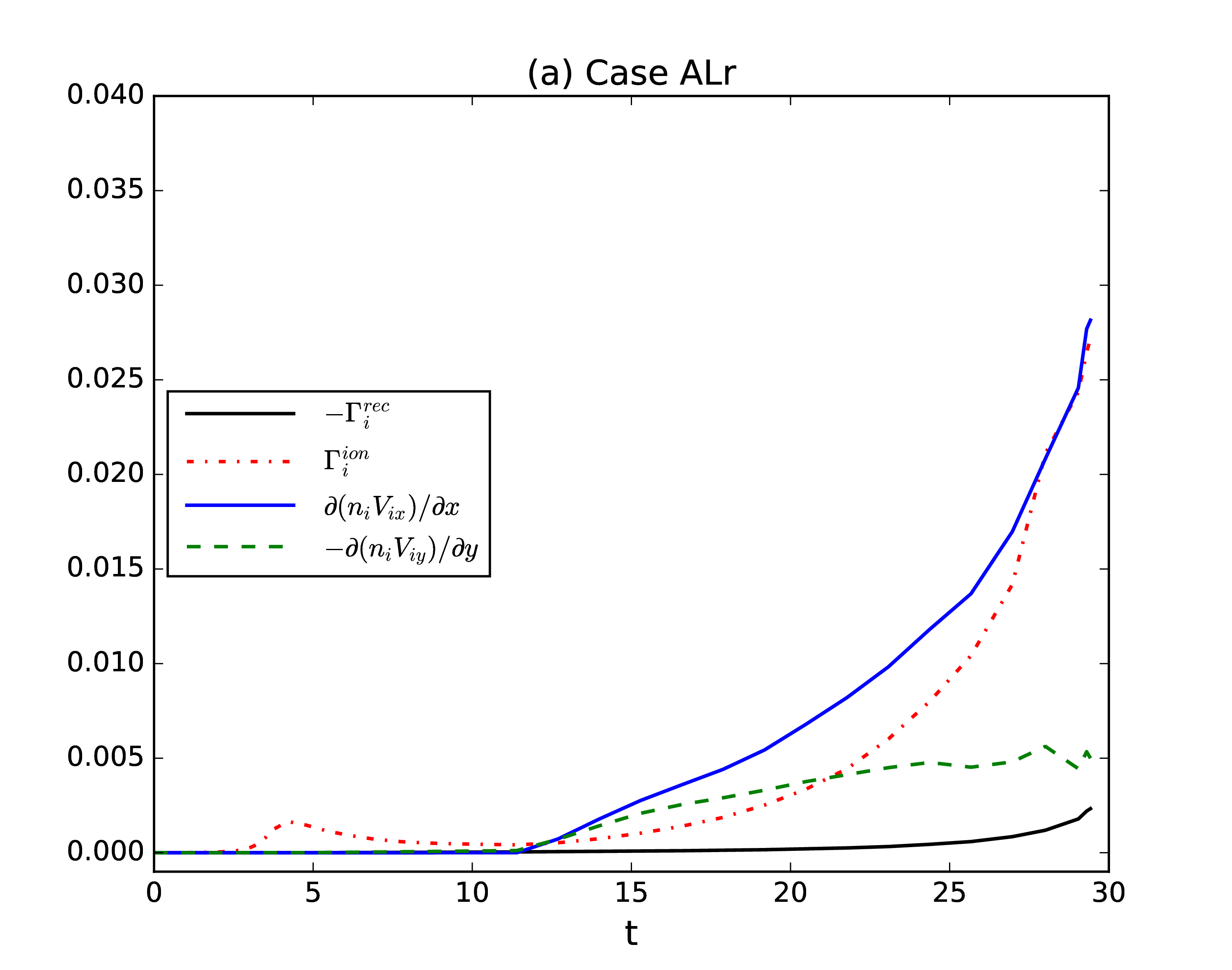}
                          \includegraphics[width=0.33\textwidth, clip=]{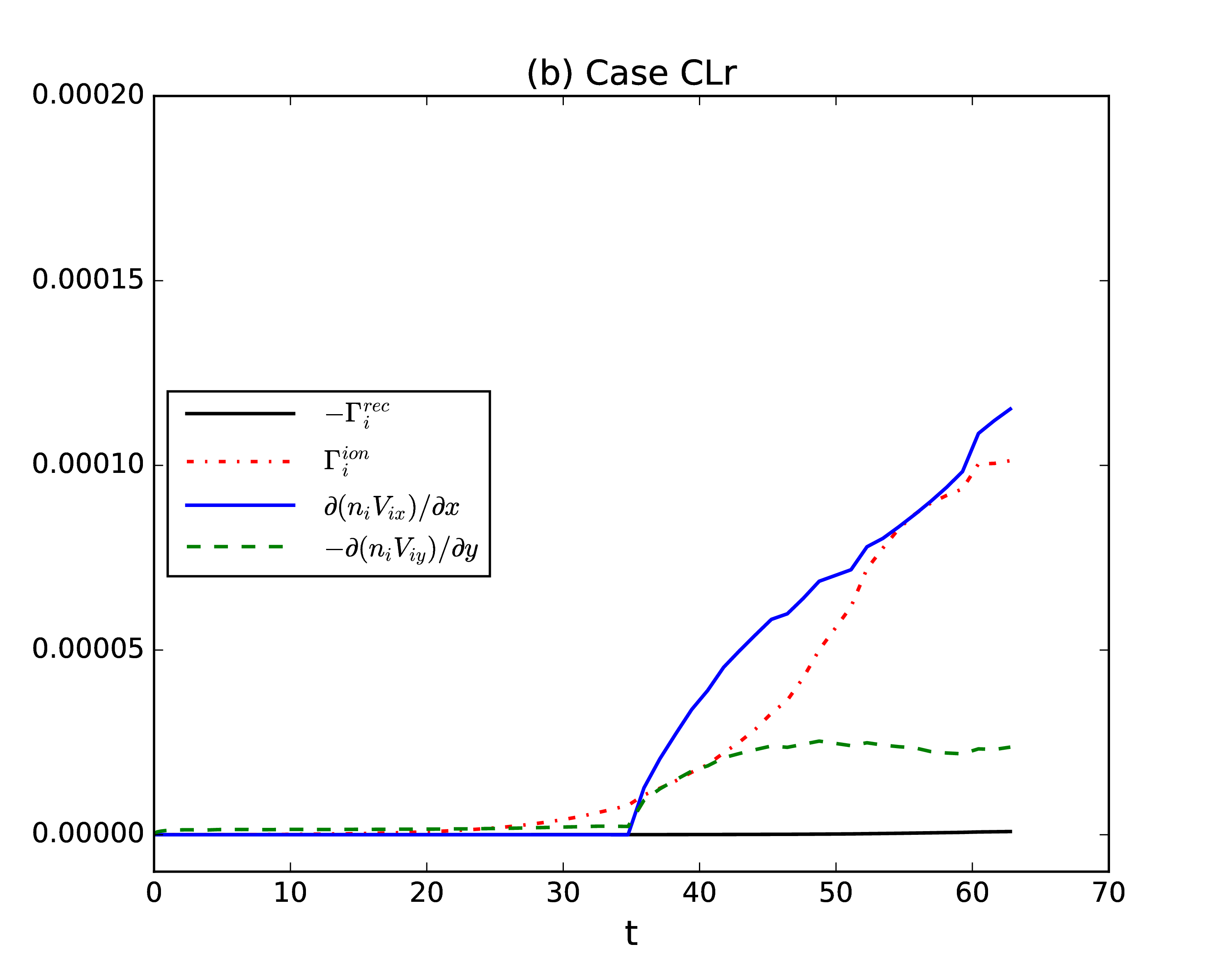}
                          \includegraphics[width=0.33\textwidth, clip=]{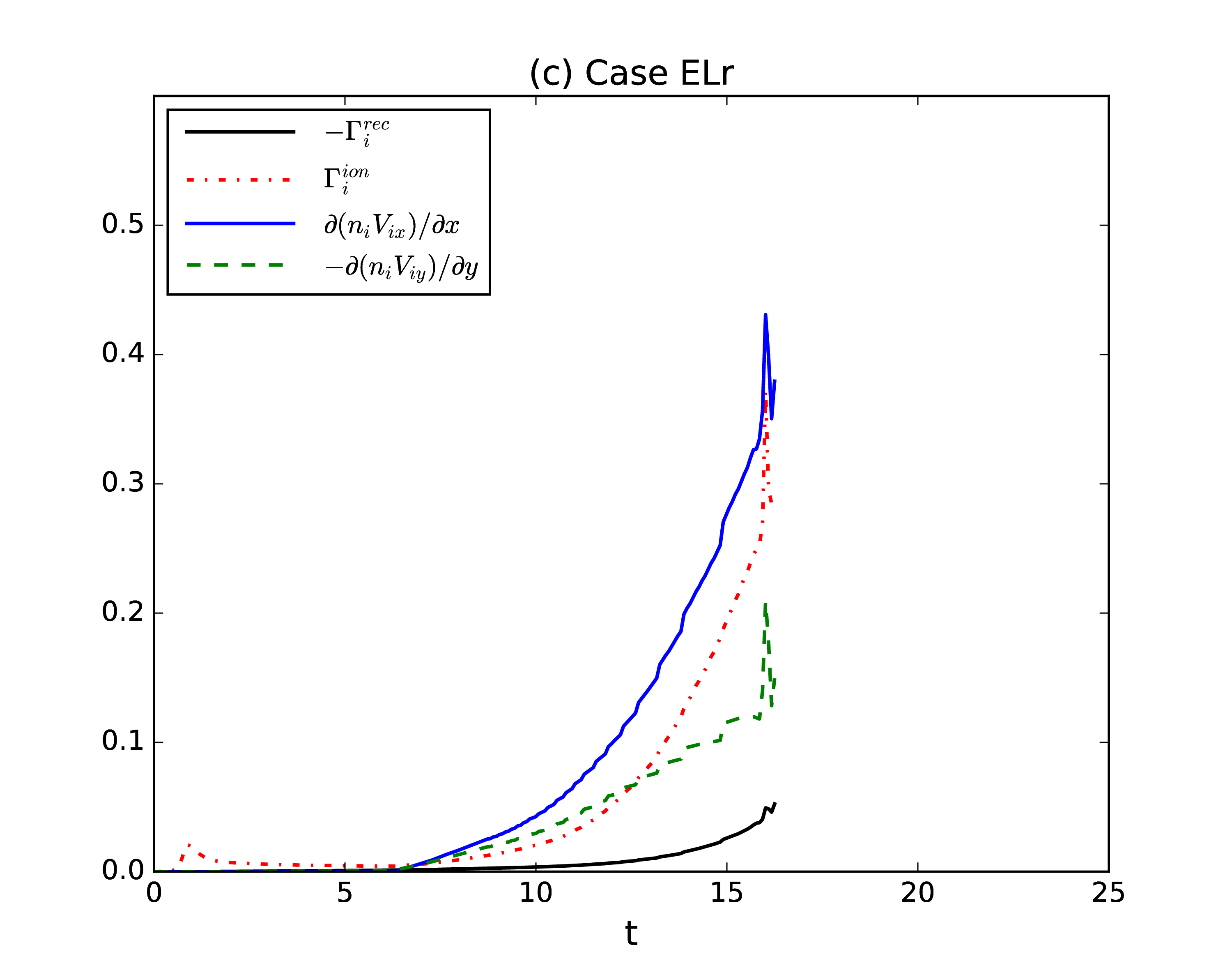} }
      \caption{Panels (a), (b) and (c) show the time dependent contributions of four components to $\partial n_i/\partial t$ in Case~ALr, CLr and ELr. The four contributions are the average values inside the current sheet, the loss due to recombination $-\Gamma_i^{rec}$;  the loss due to the outflow$\partial(n_iV_{ix})/\partial x$, the gain due to the inflow $-\partial(n_iV_{iy})/\partial y$, and the gain due to ionization $\Gamma_i^{ion}$. }
     \label{fig.4}
\end{figure}  

 \begin{figure}
      \centerline{\includegraphics[width=0.33\textwidth, clip=]{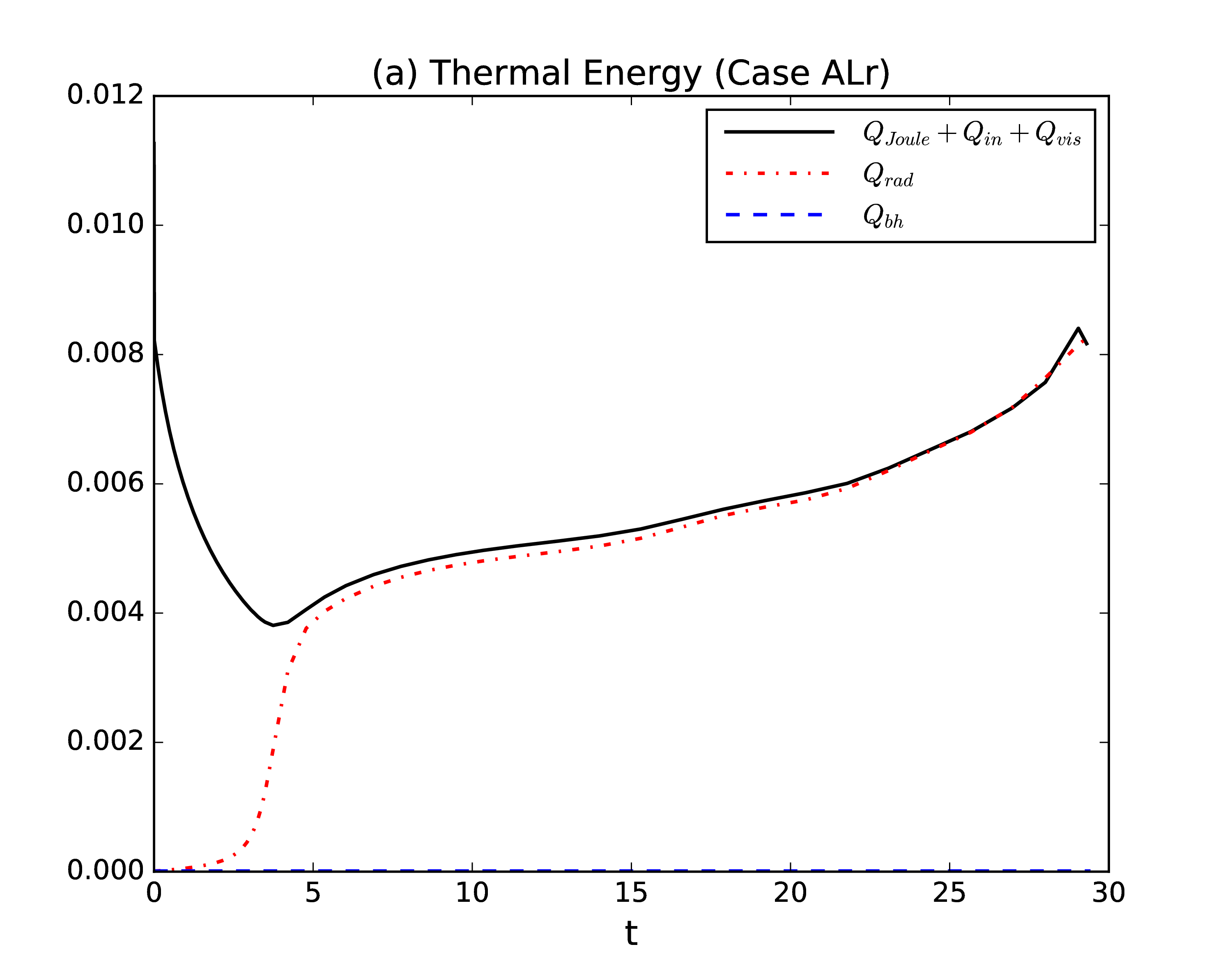}
                          \includegraphics[width=0.33\textwidth, clip=]{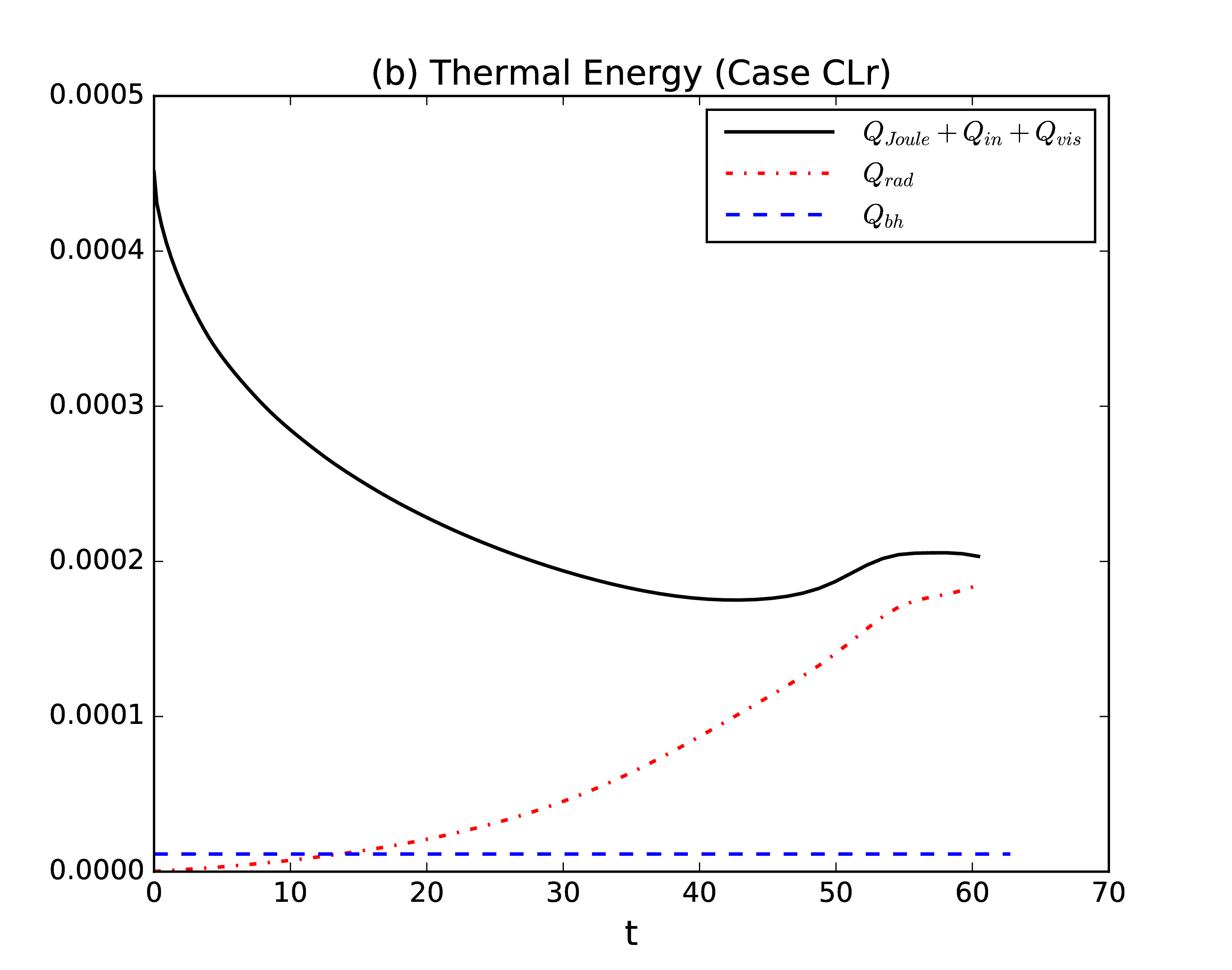}
                          \includegraphics[width=0.33\textwidth, clip=]{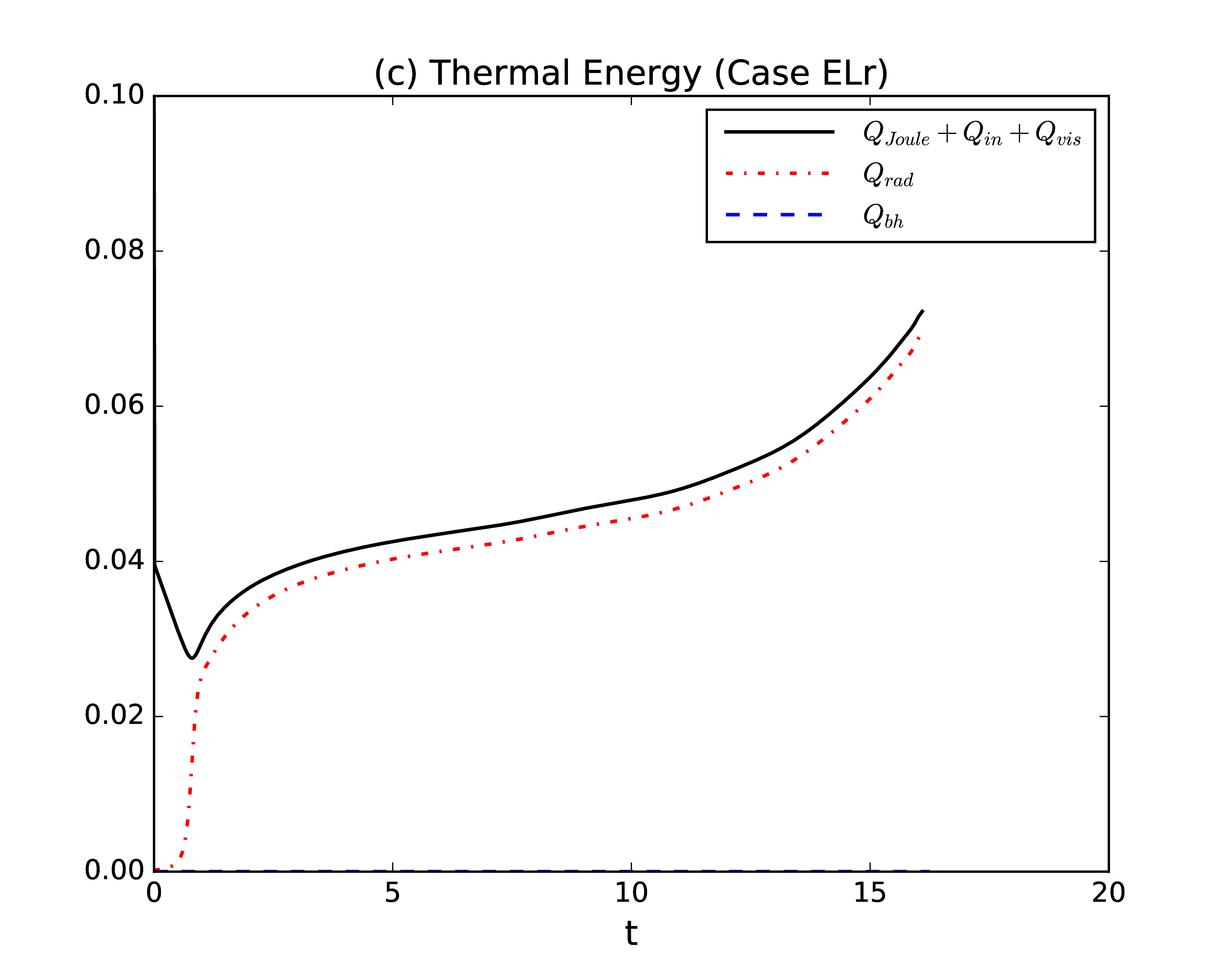} }
      \caption{(a), (b) and (c) show the time evolution of thermal energy gain and loss integrated over the simulation domain in Case ALr, CLr and ELr, respectively. The black solid lines represent the generated thermal energy $Q_{Joule}+Q_{in}+Q_{vis}$; the red dash-dotted lines represent the radiated energy $Q_{rad}$; the blue dashed lines represent the background heating $Q_{bh}$.   }
     \label{fig.5}
\end{figure}     

\begin{figure}
     \centerline{ \includegraphics[width=0.33\textwidth, clip=]{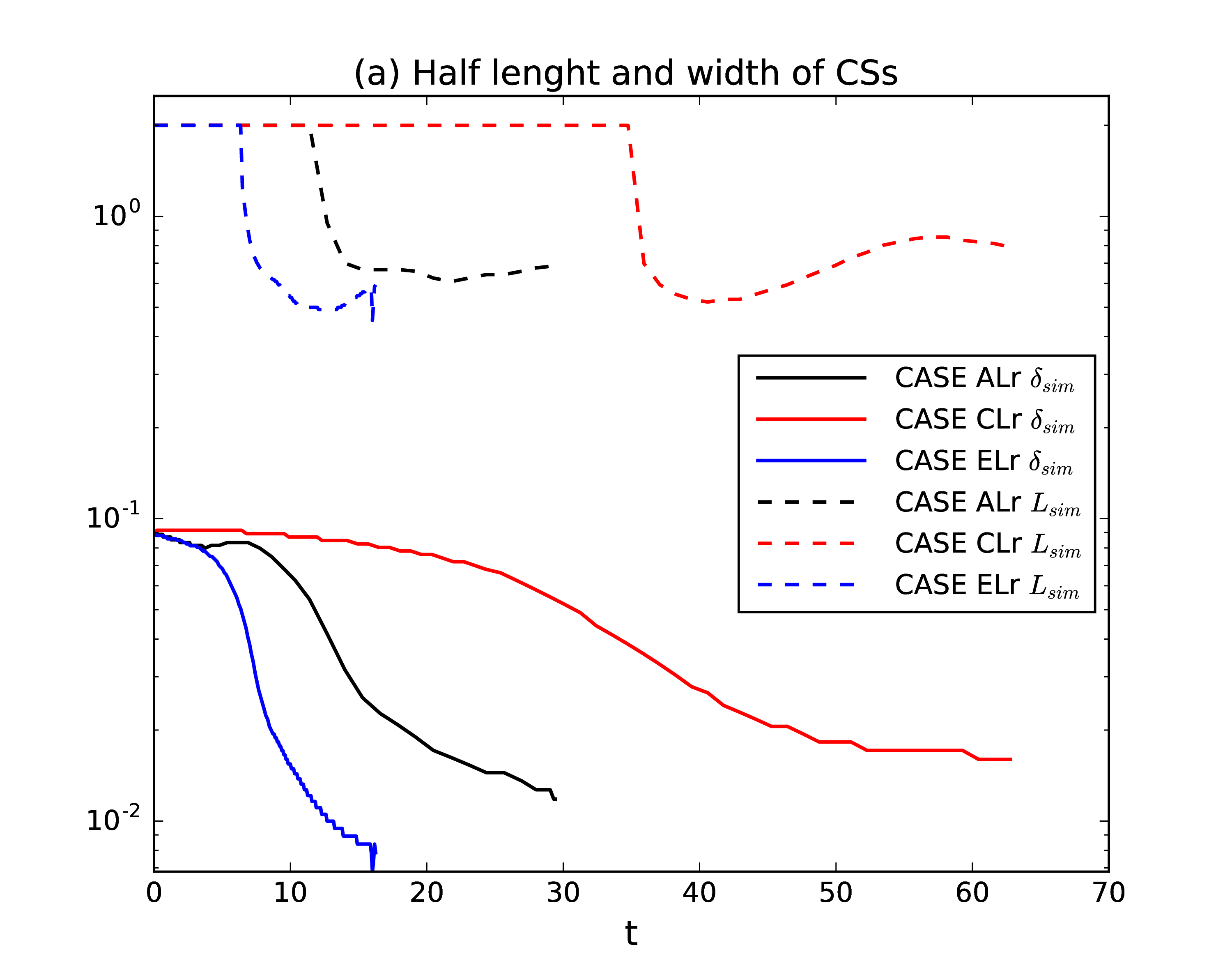}
                           \includegraphics[width=0.33\textwidth, clip=]{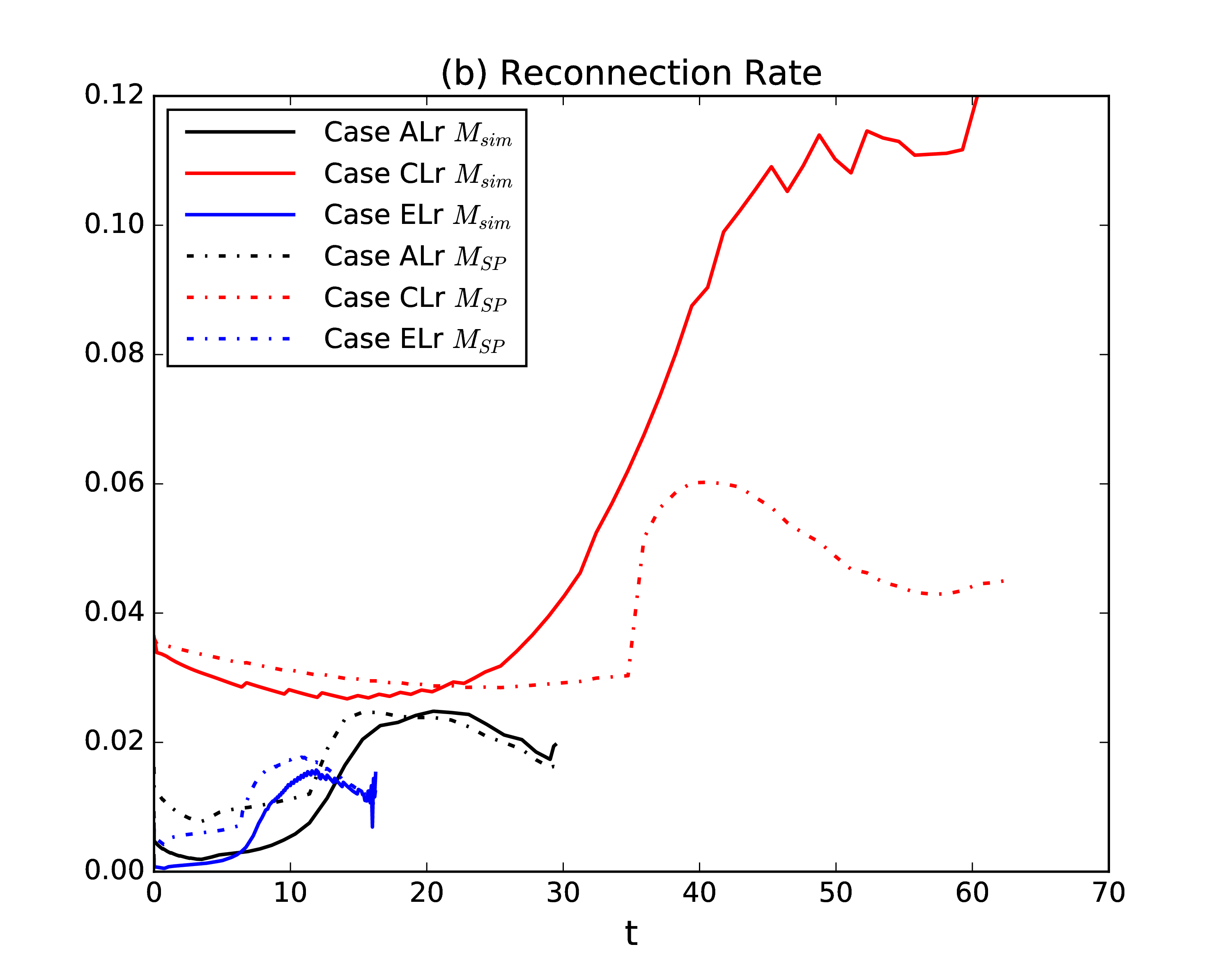} }
     \caption{(a) shows the half length $L_{sim}$ and width $\delta_{sim}$ of the current sheets in Case~ALr, CLr and ELr. (b) shows the time dependent reconnection rates in Cases~ALr, CLr, and ELr. }
     \label{fig.6}
\end{figure}

\section{Summary and discussion}\label{s:summary}
We have used the reactive multi-fluid plasma-neutral module of the HiFi modeling framework to study magnetic reconnection in strongly magnetized regions around the solar TMR, with a more realistic radiative cooling model computed using the OPACITY project and CHIANTI databases \cite{2012ApJ...751...75G}. Numerical results with different magnetic field strengths have been presented, and we have also compared the results in this work with those in our previous work \cite{2018ApJ...852...95N}  that included a simpler radiative cooling model. We summarize our results as follows:

 (1) The more realistic radiative cooling model does not result in qualitative changes of the characteristics of magnetic reconnection in strongly magnetized regions around the solar TMR. In this work, the rate of ionization of the neutral component is still faster than recombination within the current sheet region even when the initial plasma $\beta$ is as high as $\beta_0=1.46$. The ionized and neutral fluid flows are also well-coupled throughout the reconnection region for the low $\beta$ plasmas; significant decoupling of ion and neutral inflows appears in the higher $\beta$ case with $\beta_0=1.46$, which leads to a reconnection rate about three times faster than predicted by the Sweet-Parker model. The reconnection process more closely resembles the Sweet-Parker model when plasma $\beta$ is lower in our low Lundquist number simulations. 
 
 (2) In the case with stronger reconnection magnetic fields and lower plasma $\beta$, there is more thermal energy generated by Joule heating and also more radiated thermal energy in the magnetic reconnection process. The strong radiative cooling does not result in faster magnetic reconnection in strongly magnetized regions. Though most of the generated thermal energy is radiated, the maximum temperature inside the current sheet can still reach a higher value in a lower $\beta$ case. The maximum temperature is above $2\times10^4$~K when the reconnection magnetic field is higher than $500$~G. The maximum reconnection outflow velocity is above $50$~km\,s$^{-1}$ when the initial reconnection magnetic fields is as high as $1500$~G.  
 
 (3) The more realistic radiative cooling model quantitatively changes the values of some variables in the magnetic reconnection process around the solar TMR. The maximum ionization fraction is lower than that in our previous work \cite{2018ApJ...852...95N} for the same plasma $\beta$. The generated Joule heating and the radiated thermal energy in each case in this work are higher than the corresponding ones in our previous work.  
 
Our numerical results show that the ion and neutral fluids are well-coupled as a single fluid through the reconnection region in strongly magnetized regions around the solar TMR. Though most of the generated thermal energy is always dissipated by strong radiative cooling in such a high density environment, the ionization is still faster than recombination, and no acceleration of the magnetic reconnection rate is observed in the low $\beta$ environment with strong magnetic fields (above $500$~G), which is significantly different from the previous high $\beta$ simulations \cite{2012ApJ...760..109L, 2013PhPl...20f1202L, 2015ApJ...805..134M, 2017ApJ...842..117A}. The stronger reconnection magnetic fields result in the higher plasma temperature inside the current sheet. The plasma can be heated above $2\times10^4$~K when the reconnection magnetic fields are above $500$~G. Since the plasmas are still not fully ionized in Case~ELr with reconnection magnetic fields as strong as $1500$~G, the maximum temperature above $4\times10^4$~K shown in our previous work \cite{2018ApJ...852...95N} does not appear in our simulations with a stronger radiative cooling model. However, we should point out that all the simulation runs end because of the grid errors and limited resolution, not because the magnetic reconnection processes are naturally stopped. Both the ionization fraction and the plasma temperature increase with time during the reconnection process, and could possibly reach higher values if the simulations were allowed to run for a longer time. We can still expect that the plasmas may become fully ionized and higher temperature may be achieved during a magnetic reconnection process in the strongly magnetized solar TMR.

By comparing the simulation results in this work and our previous work \cite{2018ApJ...852...95N}, we can conclude that the simple radiative cooling model in our previous work is good enough for studying the characteristics of magnetic reconnection around the solar TMR with strong reconnection magnetic field ($>100$~G) and low plasma $\beta$ ($<1.46$) inside the current sheet. In such a reconnection process before the plasmas are fully ionized, the strong ionization rate $\Gamma_i^{ion}$ in the previous work allows the simple radiative cooling  $L_{rad}=\Gamma_i^{ion} \phi_{eff}$ to be comparable with the radiative cooling applied in this work for the same plasma $\beta$. However, the more realistic radiative cooling model applied in this work is necessary for studying the variations of the plasma temperature and ionization fraction inside the current sheet. It is also important for studying the energy conversion during the magnetic reconnection process when the hydrogen gas approaches full ionization. With the high temperature plasmas ($>2\times10^4$~K) likely to be generated in the Ellerman Bomb type events observed by the IRIS satellite, the hydrogen gas in the immediate neighborhood of the event is expected to become fully ionized necessitating the more realistic radiative cooling function to study the Ellerman Bomb type events and their observables.

We note that the simulation scale is only $100$~m in both this work and our previous work \cite{2018ApJ...852...95N}, which is much smaller than the observable scales of the brightening events in the solar atmosphere. Future work will show numerical results on larger scales and with much higher Lundquist numbers to reveal more physical mechanisms guiding magnetic reconnection processes and observables around the solar TMR.

\begin{acknowledgments}
The authors are grateful to the referee for the valuable comments and suggestions. This research is supported by NSFC Grants 11573064, 11203069, 11333007, 11303101 and 11403100; the Western Light of Chinese Academy of Sciences 2014; the Youth Innovation Promotion Association CAS 2017; Program 973 grant 2013CBA01503; NSFC-CAS Joint Grant U1631130; CAS grant QYZDJ-SSW-SLH012; and the Special Program for Applied Research on Super Computation of the NSFC-Guangdong Joint Fund (nsfc2015-460, nsfc2015-463, the second phase) under Grant No. U1501501. The authors gratefully acknowledge the computing time granted by the Yunnan Astronomical Observatories and the National Supercomputer Center in Guangzhou, and provided on the facilities at the Supercomputing Platform, as well as the help from all faculties of the Platform. J.L. also thanks the help from the National Supercomputing Center in Tianjin. V.S.L. acknowledges support from the US National Science Foundation (NSF). Any opinion, findings, and conclusions or recommendations expressed in this material are those of the authors and do not necessarily reflect the views of the NSF. N.A.M. acknowledges support from NSF SHINE grant AGS-135842 and DOE grant DE-SC0016363. 
\end{acknowledgments}

\end{document}